\documentclass[fleqn,usenatbib]{mnras}

\usepackage{newtxtext,newtxmath}
\usepackage[T1]{fontenc}
\usepackage{ae,aecompl}

\usepackage{bm}
\usepackage{graphicx}	% Including figure files
\usepackage{amsmath}	% Advanced maths commands
\usepackage{amssymb}	% Extra maths symbols
\usepackage{gensymb}
\title[Alignment of a circumbinary disc]{Alignment of a circumbinary disc around an eccentric binary with application to KH 15D}

\author[Smallwood et al.]{Jeremy L. Smallwood,$^1$\thanks{E-mail: Smallj2@unlv.nevada.edu}
Stephen H. Lubow,$^2$
Alessia Franchini,$^{1}$ and 
\newauthor
Rebecca G. Martin$^1$
\\
$^1$Department of Physics and Astronomy, University of Nevada, Las Vegas, 4505 South Maryland Parkway, Las Vegas, NV 89154, USA\\
$^2$Space Telescope Science Institute, Baltimore, MD 21218, USA
}

\date{Accepted XXX. Received YYY; in original form ZZZ}

\pubyear{2019}

\begin{document}
\label{firstpage}
\pagerange{\pageref{firstpage}--\pageref{lastpage}}
\maketitle

% Abstract of the paper
\begin{abstract}
We analyse the evolution of a mildly inclined circumbinary
disc that orbits an eccentric orbit binary by means of smoother particle hydrodynamic (SPH) simulations and linear theory.
We show that the alignment process of an initially misaligned circumbinary disc around an eccentric orbit binary is significantly different than around a circular orbit binary and involves tilt oscillations. The more eccentric the binary, the larger the tilt oscillations and the longer it takes to damp these oscillations.  A circumbinary disc that is only mildly inclined may increase its inclination by a factor of a few before it moves towards alignment. The results of the SPH simulations agree well with those of linear theory.  We investigate the properties of the circumbinary disc/ring around
KH 15D. We determine disc properties based on the observational constraints imposed
by the changing binary brightness. 
We find that the inclination
 is currently at a local minimum and will increase substantially
before setting to coplanarity. In addition, the nodal precession is currently near its
most rapid rate.
The recent observations
that show a reappearance of Star B impose constraints on
the thickness of the layer of obscuring material. Our results suggest that disc solids have undergone substantial inward drift and settling towards to disc midplane. For disc masses $\sim 0.001 M_\odot$, our model indicates that the level of disc turbulence is low $\alpha \ll 0.001$. Another possibility is that the disc/ring contains little gas.

\end{abstract}

% Select between one and six entries from the list of approved keywords.
% Don't make up new ones.
\begin{keywords}
accretion, accretion discs -- binaries: general -- hydrodynamics -- planets and satellites: formation
\end{keywords}

%%%%%%%%%%%%%%%%%%%%%%%%%%%%%%%%%%%%%%%%%%%%%%%%%%

%%%%%%%%%%%%%%%%% BODY OF PAPER %%%%%%%%%%%%%%%%%%

\section{Introduction}

Observations show that most stars form in relatively dense regions within stellar clusters which subsequently may be dispersed. The majority of these stars that form are members of binary star systems \citep{Duquennoy1991,Ghez1993,Duchene2013}. The observed binary orbital eccentricities vary with binary orbital period \citep{Raghavan2010, Tokovinin2016}. For short binary orbital periods, typically less than about 10 days, the eccentricities are small, likely because the orbits are circularized by stellar tidal
dissipation \citep{Zahn1977}.  The average binary eccentricity increases as a function of binary orbital period and ranges from $0.39$ to $0.59$. In addition, there is considerable scatter in eccentricity at a given orbital period with high eccentricities $\sim 0.8$ or larger sometimes found. 

Discs consisting of gas and dust likely reside within these systems at early stages. There can be multiple discs present in a binary system. A circumbinary disc orbits around the binary, while each of the binary components can be surrounded by its own disc (i.e. circumprimary and circumsecondary discs),
 as is found in binary GG Tau \citep{Dutrey1994}. Each of the discs may be misaligned to each other and to the binary. 

Some circumbinary discs have been found to be
misaligned with respect to the orbital plane of the central binary. For example, the pre-main sequence binary KH 15D has a circumbinary disc that is misaligned to the binary \citep{Chiang2004, Winn2004}. The circumbinary disc or ring around the binary protostar IRS 43 has a misalignment of at least $60\degree$ \citep{Brinch2016}, along with misaligned circumprimary and circumsecondary discs.  The binary GG Tau A may be misaligned by $25\degree$-$30\degree$ from its circumbinary disc \citep{Kohler2011,Aly2018}. There is also evidence that binary 99 Herculis, with an orbital eccentricity of $0.76$, has a misaligned debris disc that is thought to be perpendicular to the orbital plane of the binary \citep{Kennedy2012}. Furthermore, there are several known circumbinary planets discovered by Kepler, two of which have a misalignment to the binary of roughly $2.5 \degree$, Kepler-413b \citep{Kostov2014} and Kepler-453b \citep{Welsh2015}. This misalignment suggests that the circumbinary disc may have been misaligned or warped during the planet formation process \citep{Pierens2018}.

Misalignment between a circumbinary disc and the binary may occur through several possible mechanisms. First, turbulence in star-forming gas clouds can lead to  misalignment  \citep{Offner2010,Tokuda2014,Bate2012}. Secondly, if a young binary accretes material after its formation process, the accreted material is likely to be misaligned to the orbital binary plane \citep{Bate2010,Bate2018}. Finally, misalignment can occur when a binary star forms within an elongated cloud whose axes are misaligned with respect to the cloud rotation axis \cite[e.g.][]{Bonnell1992}. 

The torque from binary star systems can impact the planet formation process compared to discs around single stars \citep{Nelson2000,Mayer2005,Boss2006, Martin2014,Fu2015, Fu2015b,Fu2017}. By understanding the structure and evolution of these discs, we can shed light on the observed characteristics of exoplanets. 

Dissipation in a misaligned circumbinary disc causes tilt evolution. A disc around a circular orbit binary  aligns to the orbital plane of the binary \cite[e.g.][]{PT1995, Lubow2000, Nixonetal2011b,Facchinietal2013,Foucart2014}. However, for a disc around an eccentric binary, its angular momentum aligns to one of two possible orientations: alignment to the angular momentum of the binary orbit or, for sufficiently high initial inclination,  alignment to the eccentricity vector of the binary \citep{Aly2015,Martin2017,Lubow2018,Zanazzi2018}.
The latter state is the so-called polar configuration in which the disc plane lies perpendicular to the binary orbital plane. The timescale for the polar alignment process may be shorter or longer than the lifetime of the disc depending upon the properties of the binary and the disc \citep{Martin2018}.
%The dynamical nature of misaligned circumbinary discs have been previously studied, but these studies assumed either the binary orbit is circular \citep{Nixon2011,Facchini2013,Foucart2013,Lodato2013} or that the binary orbit is coplanar to the disc \citep{Artymowicz1994,Dunhill2015,Fleming2017}. 
%In this work, we focus on the wave-like regime in which $H/r > \alpha$, which is relevant for protoplanetary discs, and we focus on an eccentric binary that is non-coplanar to the disc.

Through SPH simulations \cite{Martin2017} found that an initially misaligned ($i = 60\degree$) low mass  circumbinary disc  around an eccentric ($e_{\rm b} =0.5$) binary undergoes damped nodal oscillations and eventually evolves to a polar configuration. \cite{Martin2018} explored the properties of binaries and discs that lead to a final polar configuration.
1D linear models for the evolution of a low mass, nearly polar disc around an eccentric binary also show evolution to a polar configuration
 \citep{Zanazzi2018,Lubow2018}.
%Planets formed in such a disc would then reside on polar orbits. Such planets would be harder to detect than the nearly coplanar planets found by recurrent transits with Kepler. Polar planets
%may be detectable as nonrecurring  transits of the binary or eclipse timing variations of the binary.  

In this paper, we extend the work of \cite{Martin2017} and 
\cite{Lubow2018}
by studying the evolution of misaligned circumbinary discs 
around eccentric orbit binaries with lower initial inclinations that ultimately result in coplanar
alignment with the binary.  We apply both 3D SPH simulations 
and 1D linear equations for a variety of disc and binary properties. 
%By mapping the structure and evolution of discs with a variety of initial conditions, we can explore the dynamics of disc alignment and the likely orientation of planetary orbits.

First we examine test particle orbits around a circular and eccentric binary in Section~\ref{test}. In Section~\ref{disc}, we use three dimensional hydrodynamical simulations of circumbinary discs to explore the evolution of aligning circumbinary discs for various values of inclination, eccentricity, and disc size. In Section~\ref{sec:model}, we apply a 1D linear model for the disc evolution. In Section~\ref{sec:expansion}, we apply the nearly rigid disc expansion procedure.
We apply our results to the observed circumbinary disc in KH 15D in Section~\ref{kh15d}.  %In Section~\ref{discussion} we discuss.... Finally, we draw our conclusions in Sections~\ref{concs}.
Section~\ref{concs} contains a summary.

\section{Test particle orbits}
\label{test}

\begin{figure*}
\includegraphics[width=6.0cm]{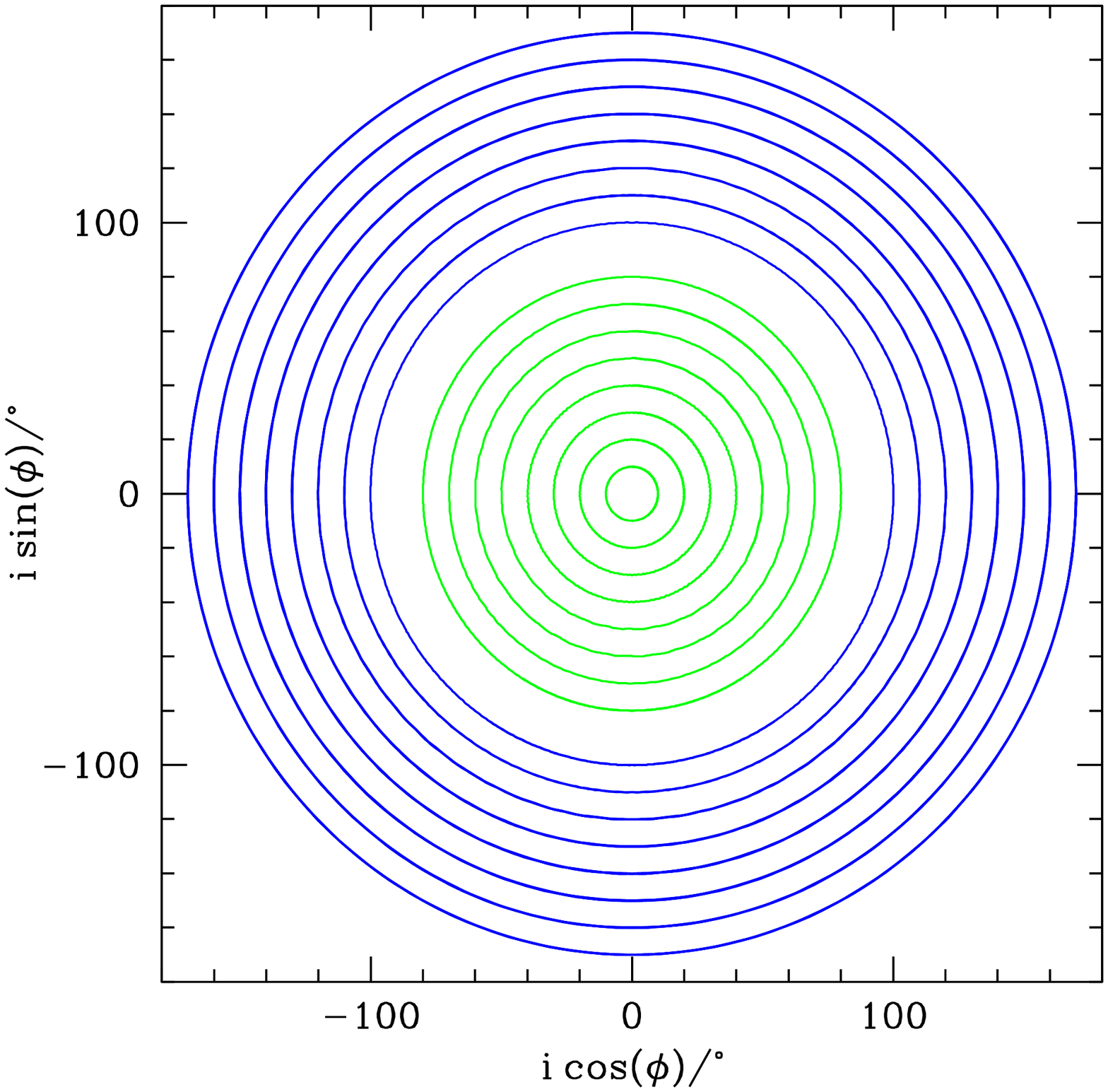}
\includegraphics[width=6.0cm]{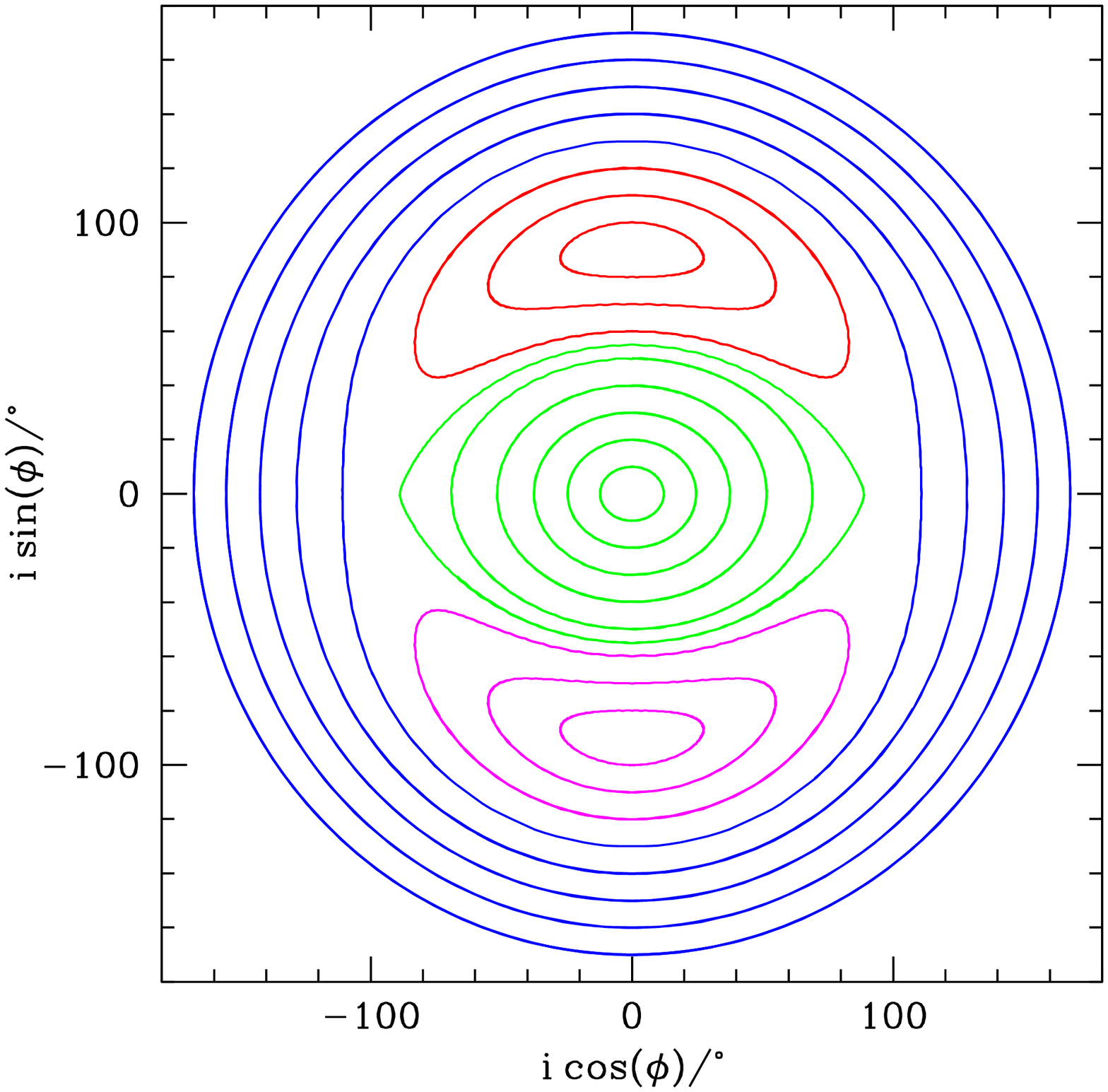}
\includegraphics[width=6.0cm]{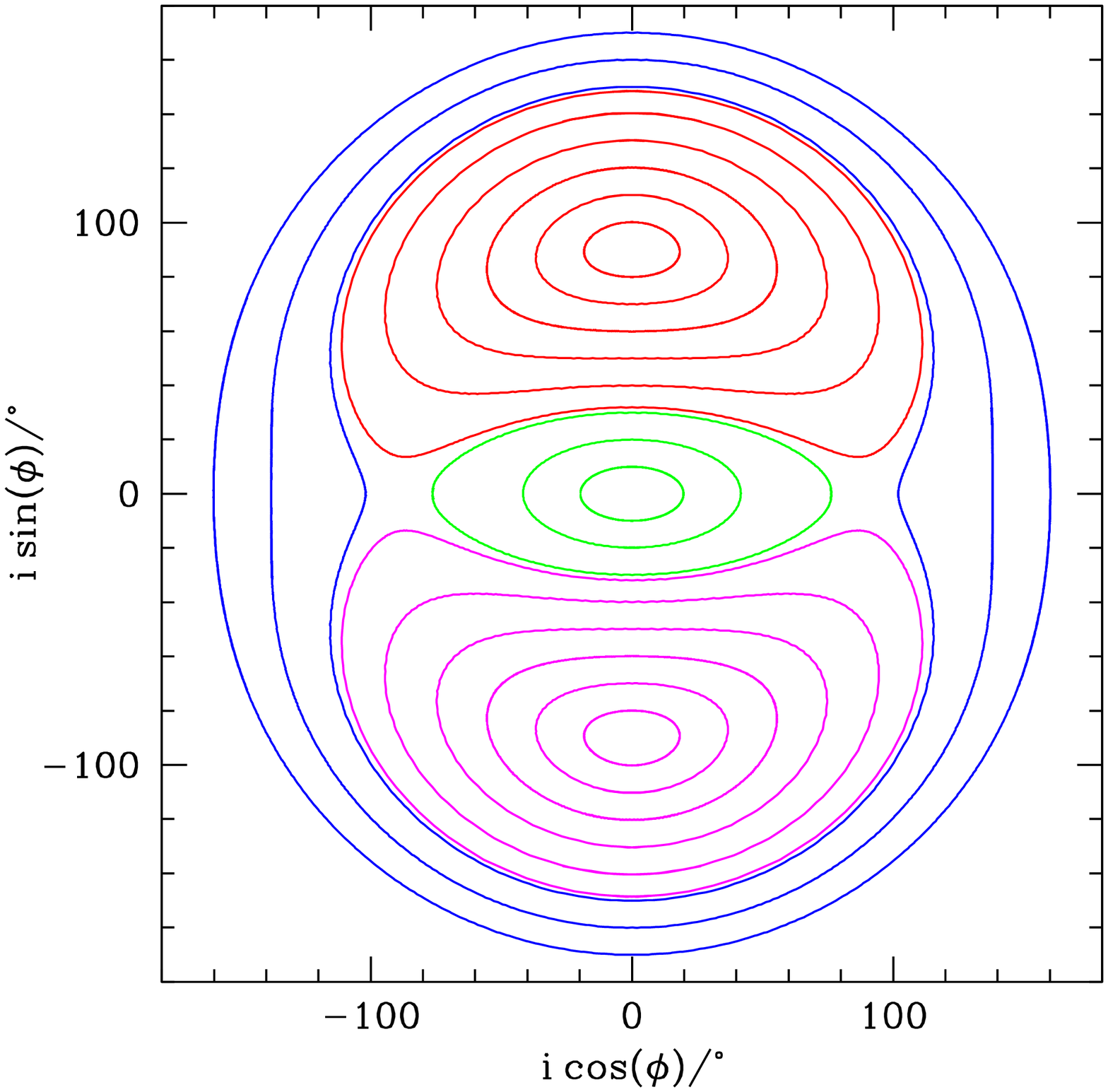}\centering
\includegraphics[width=6.0cm]{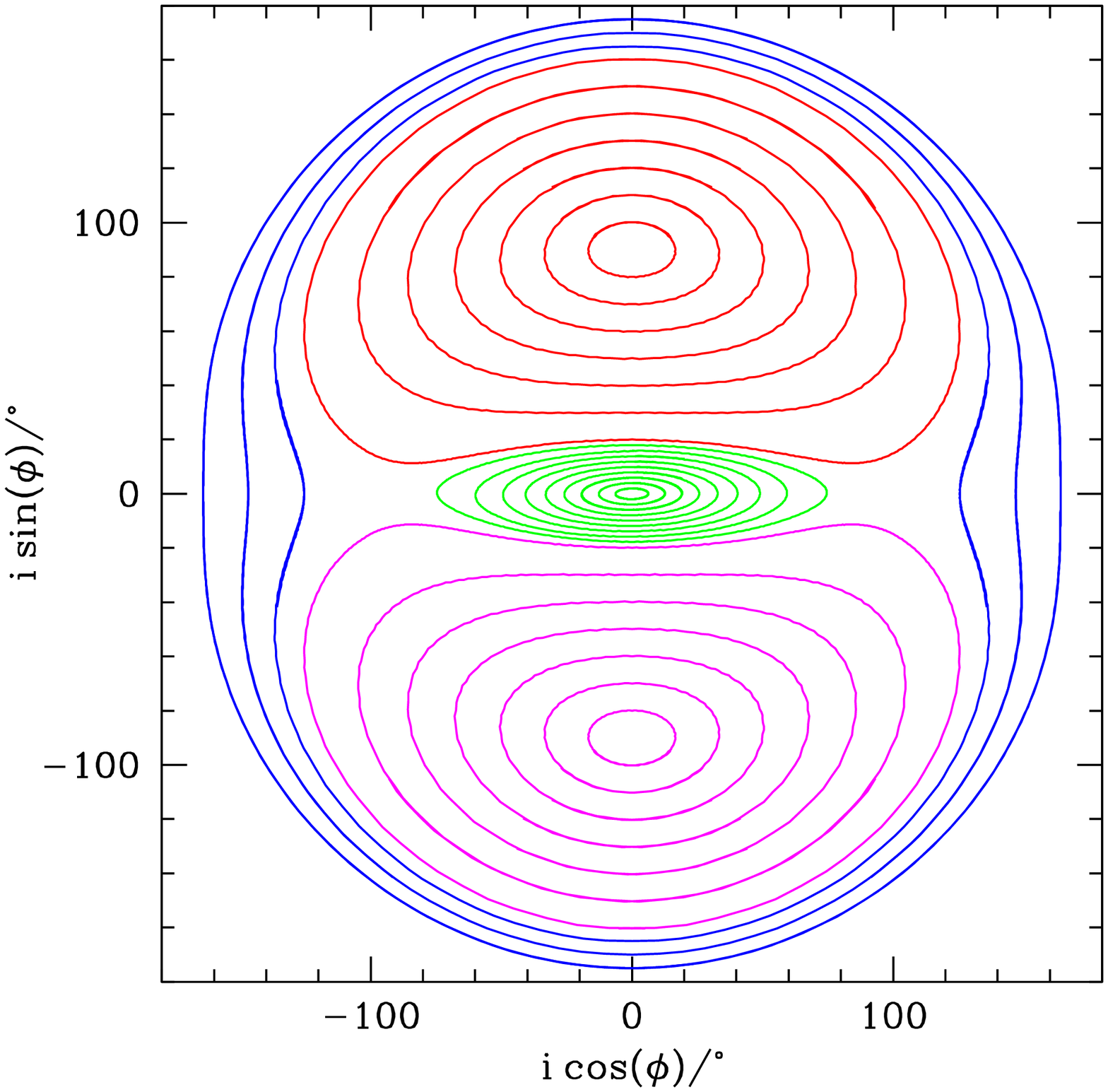} \centering
\centering
\caption{The $i\cos\phi$-$i\sin\phi$ plane for misaligned test circular particle orbits with varying initial inclination and longitude of the ascending node. The green lines show orbits close to prograde, red/magenta lines show orbits that have a librating solution and the blue lines show orbits close to retrograde. Upper left panel: A circular binary with $e_{\rm b} =0.0$. Upper right panel:  $e_{\rm b} =0.3$. Lower left panel: $e_{\rm b} =0.6$. Lower right panel:  $e_{\rm b} =0.8$. }
\label{testpart}
\end{figure*}

In this section we consider the evolution of the orbit of an inclined test particle around a binary. For a circular orbit binary, or for a sufficiently low inclination test particle orbit around an eccentric binary, the test particle orbital angular momentum precesses about the binary angular momentum. An eccentric orbit binary generates a secular potential that is nonaxisymmetric with respect to the direction of the binary angular momentum. Consequently, the  particle orbit tilt $i$ oscillates, the precession rate is nonuniform, and the precession is fully circulating.
For higher inclination around an eccentric binary, the orbit precesses about the eccentricity vector of the binary and  also undergoes oscillations in tilt. The particle in that case undergoes libration, rather than circulation \citep{Verrier2009,Farago2010,Doolin2011}.  

We consider test particle orbits around an equal mass binary with  $M_1 = M_2 = 0.5M$, where $M$ is the mass of the binary and the semi-major axis of the binary is denoted as $a$. The particle orbits are calculated for four different binary eccentricities, $e_{\rm b} =0.0$, $0.3$, $0.6$, and $0.8$. The orbital period of the binary is given by $P_{\rm orb} = 2\pi / \sqrt{G(M_1 + M_2)/a^3}$. The binary begins at periastron separation. We apply a Cartesian coordinate system $(x, y, z)$. The $x-$axis is along the binary eccentricity vector, whose direction is from the binary center of mass to the orbital pericenter. The $z-$axis is along the  binary angular momentum.
% * <lubow@stsci.edu> 2018-07-10T20:24:06.366Z:
%
% ^.
The test particle begins in a circular Keplerian orbit at position $(0,d,0)$ with velocity $(-\Omega_{\rm p} d \cos i_0,0,\Omega_{\rm p} d \sin i_0)$ where $\Omega_{\rm p} = \sqrt{G(M_1+M_2)/d^3}$ is approximate angular frequency of a particle about  the center of mass of the binary and $i_0$ is the initial particle orbit tilt with respect to the binary orbital plane. The longitude of the ascending node $\phi$ is measured from the $x$-axis. These initial conditions correspond to an initial longitude of the ascending node of $\phi_0 = 90\degree$.

Fig.~\ref{testpart} shows the test particle orbits in the $i\cos\phi$-$i\sin\phi$ phase space for binary eccentricities of $e_{\rm b}  = 0.0$ (upper left panel), $0.3$ (upper right panel), $0.6$ (lower left panel), and $0.8$ (lower right panel) for various initial inclinations. The test particles all begin at a separation of $d=5a$. For these test particle orbits, the separation does not affect these phase portraits, only the timescale on which the orbit precesses. Depending on the initial orbital inclination, the particle can reside on a circulating or librating orbit. The centers of the upper libration regions (for all panels except the circular case) corresponds to $i = 90\degree$ and $\phi = 90\degree$, while the centers for the lower librating regions correspond to $i = 90\degree$ and $\phi = -90\degree$. 

For higher binary eccentricity, the critical inclination angle that separates the librating solutions from circulating solutions is smaller. When the third body (in this case a test particle) is massive, the nodal libration regions shrink \citep[see Fig.~5 in][]{Farago2010}.  The critical inclination for test particles that divides the librating and circulating solutions is
\begin{equation}
i_{\rm crit}=\sin^{-1}\sqrt{\frac{1-e_{\rm b}^2}{1+4e_{\rm b}^2}}
\label{icrit}
\end{equation}
\citep{Farago2010}. For the eccentricities considered in Fig.~\ref{testpart} this is $i_{\rm crit}=54.9^\circ$ for $e_{\rm b} =0.3$,  $i_{\rm crit}=30.8^\circ$ for $e_{\rm b} =0.6$ and  $i_{\rm crit}=18.5^\circ$ for $e_{\rm b} =0.8$.  \cite{Martin2018} found that the critical inclination is slightly higher for a disc than a test particle. This means that a disc is more likely to move towards coplanar alignment with the binary than a test particle. In the next section we consider the evolution of a hydrodynamic circumbinary disc and use these test particle orbits for comparison.

\begin{table}
	\centering
	\caption{Parameters of the initial circumbinary disc around an equal mass binary with total mass $M$, and separation $a$.}
%	\label{tab:example_table}
	\begin{tabular}{lll} % four columns, alignment for each
		\hline
	    Binary and Disc Parameters & Symbol & Value\\
		\hline
		Mass of each binary component & $M_1/M = M_2/M$& $0.5$ \\
		Accretion radius of the masses & $r_{\rm acc}/a$ & $0.25$ \\
        Initial disk mass & $M_{\rm di}/M$ & $0.001$\\
     	Initial disk inner radius & $r_{\rm in}/a$  & $2$\\
        Disc viscosity parameter & $\alpha$  & 0.01\\
        Disc aspect ratio & $H/r (r = r_{\rm in})$  & 0.1\\
		\hline
	\end{tabular}
    \label{table1}
\end{table}

\begin{table}
	\centering
	\caption{The setup of the SPH simulations which lists the eccentricity of the binary, $e_{\rm b}$, the initial tilt of the disc, $i_0$, and the initial outer boundary of the disc, $r_{\rm out}$. We also list the critical inclination of a test particle derived from Equation~(\ref{icrit}). The initial tilts from each model are always below the critical to assure the disc aligns to the orbital binary plane. }
%	\label{tab:example_table}
	\begin{tabular}{lllll} % four columns, alignment for each
		\hline
	    Model & $e_{\rm b}$ & $i_0$ & $i_{\rm crit}$ & $r_{\rm out}/a$\\
		\hline
		Run1 & $0.0$ & $60\degree$ & -- & $5$ \\
		Run2 & $0.3$ & $50\degree$ & $54.9\degree$ & $5$ \\
        Run3 & $0.6$ & $30\degree$ & $30.8\degree$ & $5$ \\
     	Run4 & $0.8$ & $15\degree$ & $18.5\degree$ & $5$ \\
        Run5 & $0.8$ & $15\degree$ & $18.5\degree$ & $40$ \\
        Run6 & $0.3$ & $10\degree$ & $54.9\degree$ & $5$ \\
        Run7 & $0.6$ & $10\degree$ & $30.8\degree$ & $5$ \\
        Run8 & $0.8$ & $10\degree$ & $18.5\degree$ & $5$ \\
		\hline
	\end{tabular}
    \label{table2}
\end{table}

\section{Circumbinary Disc Simulations}
\label{disc}
To model the alignment process of misaligned circumbinary discs around an eccentric binary, we use the 3D smoothed particle hydrodynamics \cite[SPH; e.g.,][]{Price2012a} code {\sc phantom} \citep{Lodato2010,Price2010,price2017}. {\sc phantom} has been well tested and used to model misaligned accretion discs in binary systems \citep{Nixon2012,Nixon2013,Martin2014,Dougan2015}.

\begin{figure*} 
\includegraphics[width=14cm]{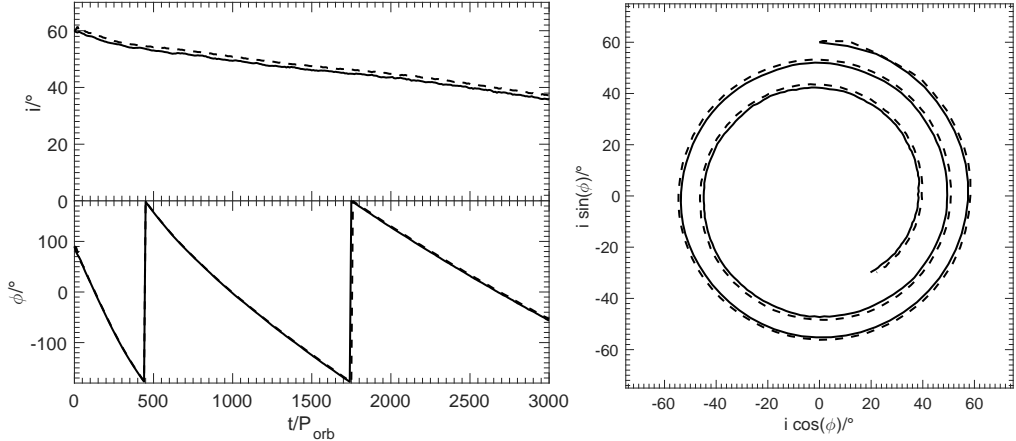}
\centering
\caption{Left: The upper panel shows the inclination, $i$, and lower panel the longitude of the ascending node, $\phi$, for a circumbinary disk with $i_0 = 60\degree$ around a circular binary, $e_{\rm b} = 0.0$, (Run1). Right: the $i\cos\phi$-$i\sin\phi$ phase space. The measurements are taken within the disc at a distance of $3a$ (solid) and $5a$ (dashed).}
\label{e00}
\end{figure*}

\subsection{Simulation Setup}
\label{disc_setup}

Table~\ref{table1} summarises the initial conditions of the binary and disc parameters for the hydrodynamical simulations.  We consider an equal mass binary with total mass $M=M_1 + M_2$. The eccentric orbit of the binary lies in the $x$-$y$ plane with semi--major axis, $a$.  The binary begins at time $t=0$ at apastron. The accretion radius of each binary component is $0.25a$. When a particle enters this radius, it is considered accreted and the particle's mass and angular momentum are added to the sink particle. We consider binaries with eccentricities $e_{\rm b}  = 0.0$, $0.3$, $0.6$ and $0.8$. For each eccentricity, we begin with a low initial disc inclination somewhat below the critical value found from Equation~(\ref{icrit}). Table~\ref{table2} summarises the setup for each simulation. For $e_{\rm b} =0.3$ we use $i=50\degree$, for $e_{\rm b} =0.6$ we use $i = 30\degree$ and for $e_{\rm b} =0.8$ we use $i = 15\degree$ We evolve each simulation to $3000$ binary orbits. 

Each simulation has an initially low disc mass of $10^{-3}\, M$ and we ignore self--gravity. The low mass disc has a negligible dynamical affect on the orbit of the binary. Each simulation consists of $6 \times 10^5$ equal mass gas particles that initially  reside in a flat disc with an inner boundary of $2a$ and an outer boundary of $5a$. The inner boundary of the disc is chosen to be close to where the tidal torque truncates the inner edge of the disc \citep{Artymowicz1994}. For misaligned discs, the tidal torque produced by the binary is much weaker allowing the disc to move closer to the binary \cite[e.g.,][]{Lubow2015,Miranda2015,Nixon2015, Lubow2018}.  The surface density profile is initially a power law distribution $\Sigma \propto R^{-3/2}$. We use a locally isothermal disc with sound speed $c_{\rm s} \propto R^{-3/4}$ and disc aspect ratio $H/r = 0.1$ at $r = r_{\rm in}$. We take the \cite{shakura1973} $\alpha$ to be $0.01$. From these values we derive an artificial viscosity ($\alpha_{\rm AV}$) of $0.4$ (a value of $\alpha_{\rm AV} = 0.1$ represents the lower limit, below which a physical viscosity is not resolved in SPH) and set $\beta_{\rm AV} = 2.0$ from the SPH description detailed in \cite{Lodato2010} which is given as
\begin{equation}
\alpha \approx \frac{\alpha_{\rm AV}}{10}\frac{\langle h \rangle}{H},
\end{equation}
where $\langle h \rangle$ is the mean smoothing length on particles in a cylindrical ring at a given radius \citep{price2017}. With this value of $\alpha$, the disc with an initial outer radius of $5a$ is resolved with a shell-averaged smoothing length per scale height of $\langle h \rangle/H \approx 0.25$. For the simulation with a larger outer radius of $40a$, we have that $\langle h \rangle/H \approx 0.30$.

%\begin{figure*} \centering
%\includegraphics[width=17cm]{new_e0p8_i70_aug90.eps}
%\includegraphics[width=17cm]{new_e0p8_i80_aug90.eps}
%\includegraphics[width=17cm]{new_e0p8_i90_aug90.eps}\centering
%\caption{ 1) [e=0.8 i = 70 aug=90] 2) [e=0.8 i = 80 aug=90] 3) [e=0.8 i = 90 aug=90] 4) [e=0.8 i = 100 aug=90]  600,000 particles  }
%\label{}
%\end{figure*}

\begin{figure*} 
\includegraphics[width=14cm]{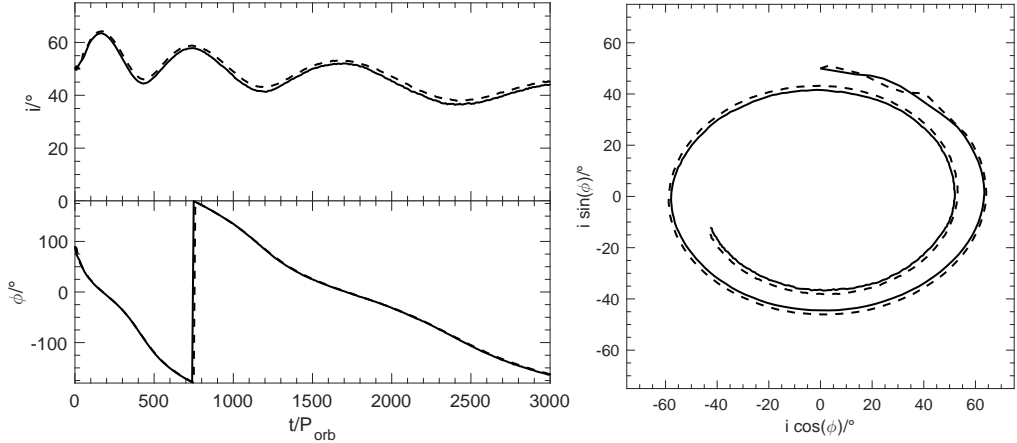}
\centering
\caption{Same as Fig.~\ref{e00} but for a circumbinary disk with $i_0 = 50\degree$ and binary eccentricity $e_{\rm b} = 0.3$ (Run2).}
\label{e03}
\end{figure*}

\begin{figure*} \centering
\includegraphics[width=14cm]{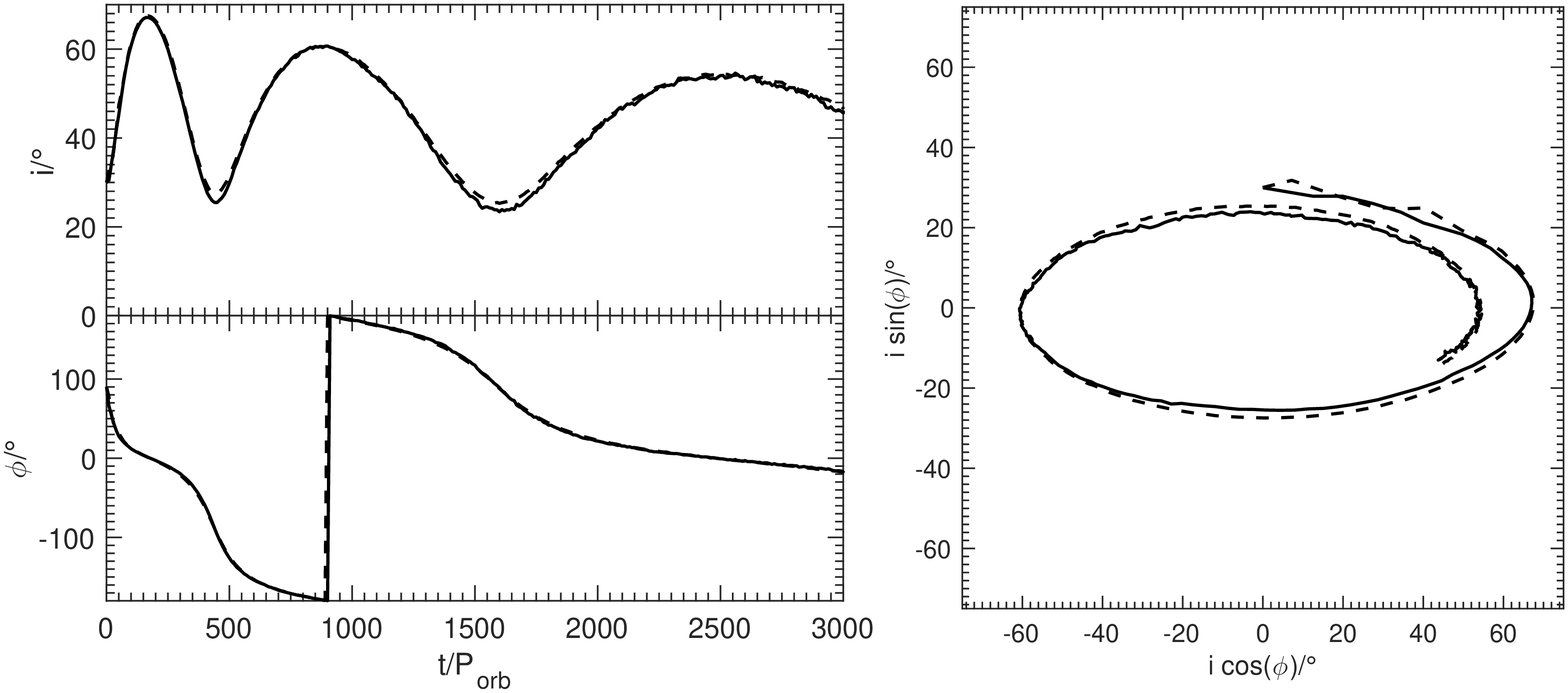}\centering
\caption{Same as Fig.~\ref{e00} but for a circumbinary disk with $i_0 = 30\degree$ and binary eccentricity $e_{\rm b} = 0.6$ (Run3).}
\label{e06}
\end{figure*}

\begin{figure*} \centering
\includegraphics[width=17.6cm]{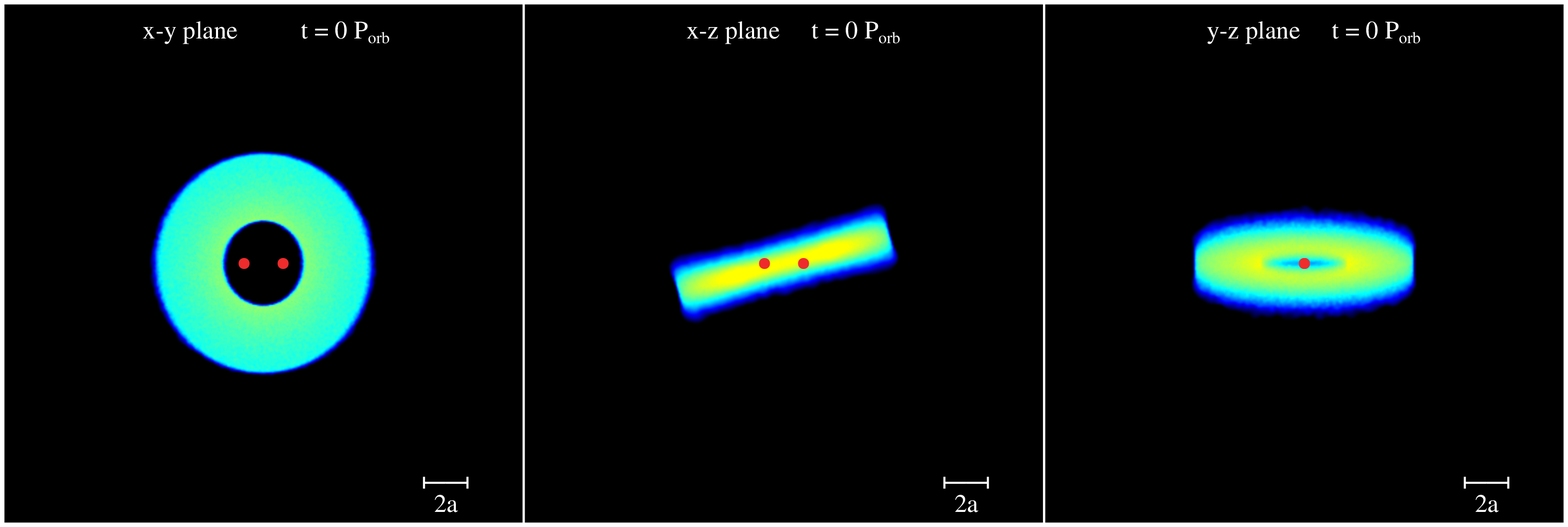}
\includegraphics[width=17.6cm]{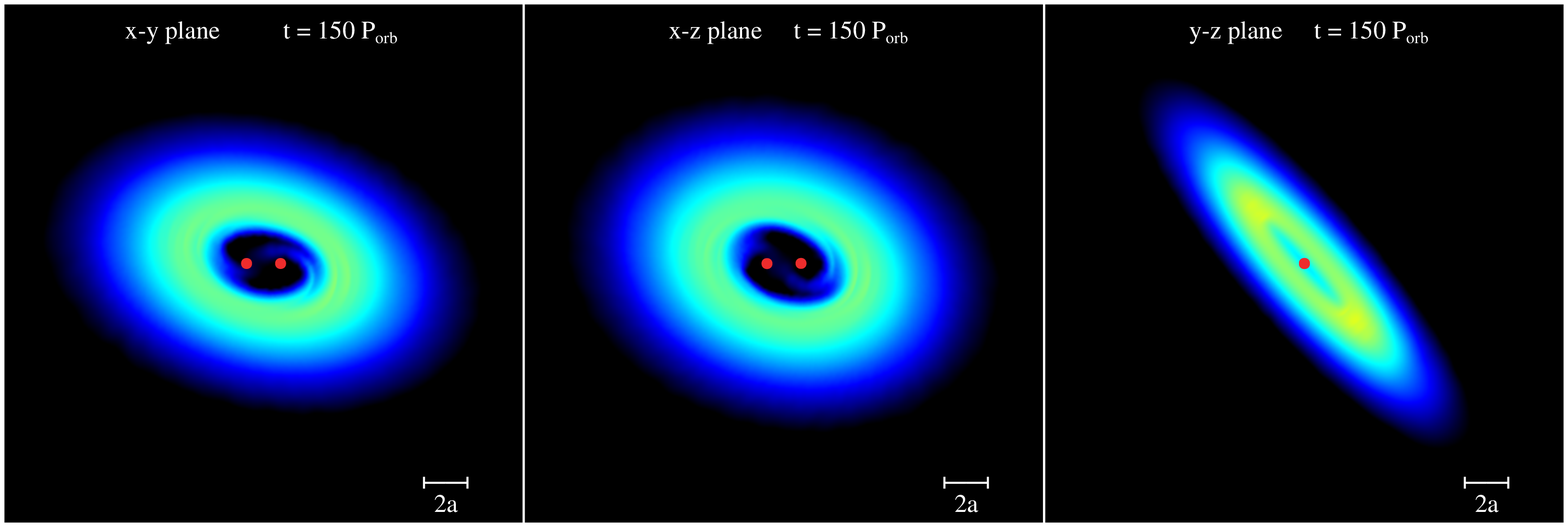}\centering
\caption{Disc evolution for a circumbinary disk with $i_0 = 15\degree$ around a binary with $e_{\rm b} = 0.8$ (Run4). Upper panels: initial disc setup for the {\sc phantom} SPH simulation of an eccentric binary with separation $a$ (shown by the red circles) with an inclined circumbinary disc. Lower panels: the disc at a time of $t = 150\,{\rm P_{orb}}$. The color denotes the gas density with yellow regions being about two orders of magnitude larger than the blue.  The left panels show the view looking down on to the binary orbital plane, the $x$--$y$ plane. The middle panels show the $x$--$z$ plane and the right panels show the $y$--$z$ plane. }
\label{SPH}
\end{figure*}

\begin{figure*} 
\includegraphics[width=14cm]{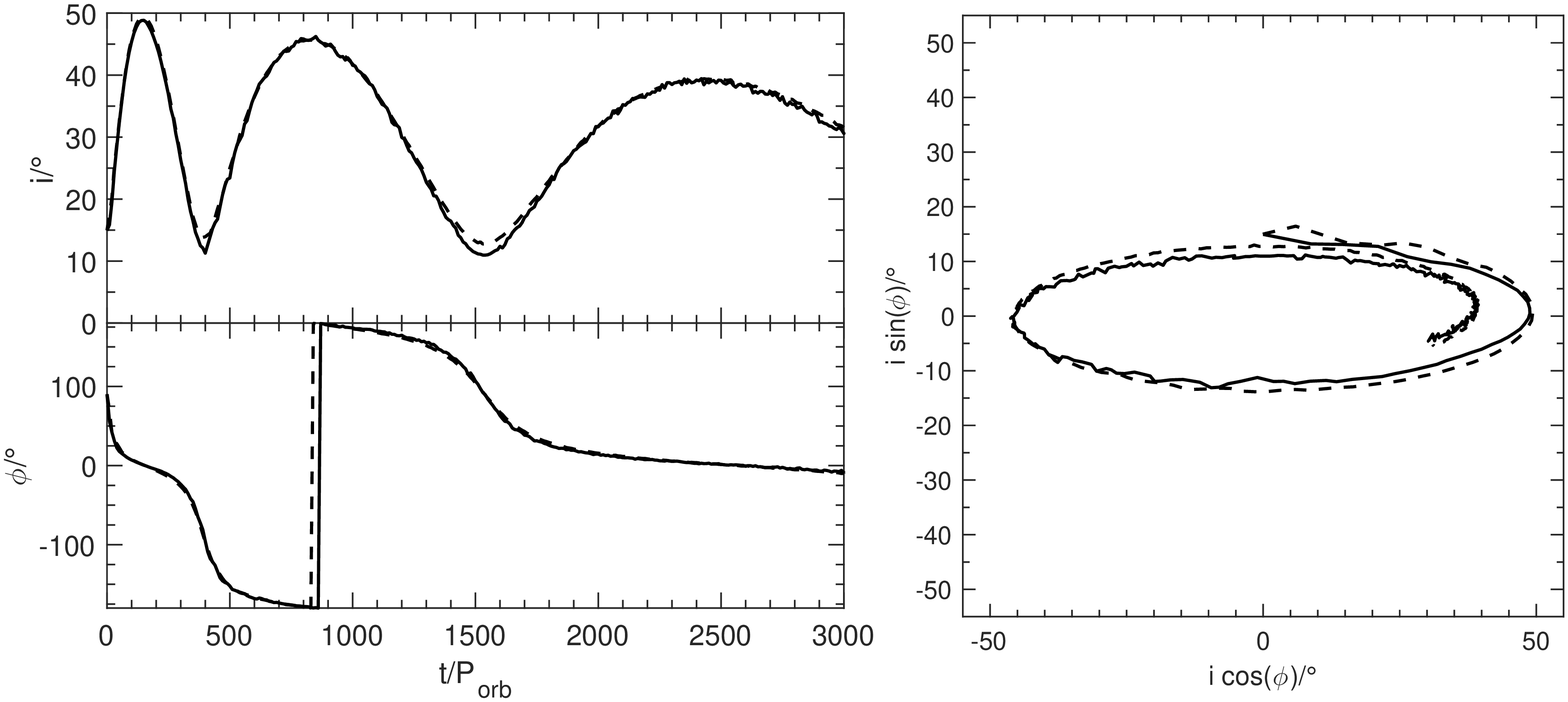}
\centering
\centering
\caption{ Same as Fig.~\ref{e00} but for a circumbinary disk with $i_0 = 15\degree$ and binary eccentricity $e_{\rm b} = 0.8$ (Run4).}
\label{e08_1}
\end{figure*}

\subsection{Results}

In this section we describe the results of the hydrodynamical disc simulations for different values of the eccentricity of the binary orbit. %For each binary eccentricity we consider, we adopt
%an initial disc inclination angle that is somewhat below the critical value given by Equation (\ref{icrit}).

\subsubsection{Circular binary  with $e_{\rm b}  = 0.0$}

The left hand panel of Fig.~\ref{e00} shows the time evolution of the inclination and longitude of ascending node at a distance $3a$ (solid lines) and $5a$ (dashed lines) of a misaligned disc with an initial inclination of  $60\degree$ around an circular binary (Run1 of Table \ref{table2}). The inclination evolution of the disc shows that the disc is aligning to the binary orbital plane. Through viscous dissipation, the disc orbital angular momentum vector evolves towards alignment with the orbital angular momentum vector of the binary.  The disc undergoes retrograde precession at a nearly constant (uniform)
precession rate about the binary angular momentum vector. The disc inclination decreases monotonically. The right hand panel shows a spiral in the $i\cos\phi$-$i\sin\phi$ phase space as the disc aligns to the binary orbital plane.

\subsubsection{Eccentric binary  with $e_{\rm b}= 0.3$}

We consider a binary eccentricity of 0.3. %, since this is the average observed binary eccentricity \citep[e.g.,][]{Duquennoy1991}.
Fig.~\ref{e03} shows the time evolution of the inclination and longitude of ascending node at a distance $3a$ and $5a$ of an initially misaligned disc of $50\degree$ around the eccentric binary (Run2 of Table \ref{table2}). The disc evolves towards alignment to the plane of the binary as in the circular binary case. However, during this process the disc undergoes tilt oscillations due to the eccentricity of the binary. The precession rate is nonuniform.

\subsubsection{Eccentric binary  with $e_{\rm b}= 0.6$}

The left hand panel of Fig.~\ref{e06} shows the time evolution of the inclination and longitude of ascending node at a distance $3a$ and $5a$ for a misaligned disc with an initial inclination of  $30\degree$ around a binary with eccentricity $e_{\rm b}=0.6$ (Run3 of Table \ref{table2}). The right hand panel shows the spiral in the $i\cos\phi$-$i\sin\phi$ phase space as the disc aligns to the binary orbital plane. The precession rate is more nonuniform than in the case
of $e_{\rm b}=0.3$ shown in Fig.~\ref{e03} and the inclination oscillations are stronger.

%The lower panels in Fig.~\ref{e06} shows the evolution of the disc with an initial misalignment of $15\degree$. The inclinations of $10\degree$ and $15\degree$ are well below the critical inclination where the disc undergoes polar alignment. This critical inclination can be estimated by the test particle orbits in Fig.~\ref{testpart}, where any inclination above this critical has a librating solution for the test particle orientation.

\subsubsection{Eccentric binary  with $e_{\rm b} = 0.8$}
 
Finally, we consider  a highly eccentric binary with $e_{\rm b}=0.8$.
This eccentricity is at the upper end of the
values for binary KH 15D determined by \cite{Johnson2004}. We consider an initial misalignments of $15\degree$ (Run4 of Table \ref{table2}). We show the initial orientation in the three Cartesian planes in the upper panels in Fig.~\ref{SPH}. In the lower panels, we show the disc orientation at a time of $t = 150\, P_{\rm orb}$ when the disc tilt has increased to about $50\degree$.  The upper left hand panel in Fig.~\ref{e08_1} shows the evolution of the tilt and the longitude of the ascending node.  The right hand panel shows the $i\cos\phi$-$i\sin\phi$ phase space plot as the disc aligns to the binary orbital plane. As expected, as the binary eccentricity increases, the amplitude of the tilt oscillations also increases as expected from the test particle orbit case. In addition, the precession rate is more nonuniform, as seen in
the lower left panel of Fig.~\ref{e08_1}.

%We also investigated the case where the circumbinary disc is initially misaligned by $20\degree$ around an eccentric binary $e_{\rm b} =0.8$. Note that for an initial inclination of $i_0 = 20\degree$ and longitude of the ascending node of $\phi_0 = 90\degree$, a test particle's orbital orientation is initially in the circulating regime but is very close to the librating regime shown in the lower right panel in Fig.~\ref{testpart}. Unlike a test particle, the disc begins in the circulating regime, but then quickly transitions into the librating regime. This means that the critical inclination for the transition between the  circulating solution and the librating solution is lower for a disc than a test particle. During this process, the disc in this case undergoes large oscillation in tilt, reaching a maximum tilt of about $160\degree$. This drastic increase in tilt causes the disc to break. {\bf Show figures to support this.} The broken precessing disc later breaks into more pieces and each precesses on a different timescale. See \cite{Nixon2013} for details on various precessing discs in the context of a binary black hole.

\subsubsection{Eccentric binary with a large disc}
The simulations described thus far only dealt with moderately extended discs with a radial extent  initially from $2a$ up to $5a$. For parameters relevant to protoplanetary discs, such  discs precess in nearly solid body because the sound crossing timescale is shorter than the precession timescales. As discussed in \cite{Martin2018}, close binaries may have a disc with a much larger radial extent relative to the binary separation. We now consider the disc evolution with a larger initial disc outer radius (Run5 of Table \ref{table2}) .

\begin{figure*} \centering
\includegraphics[width=17.6cm]{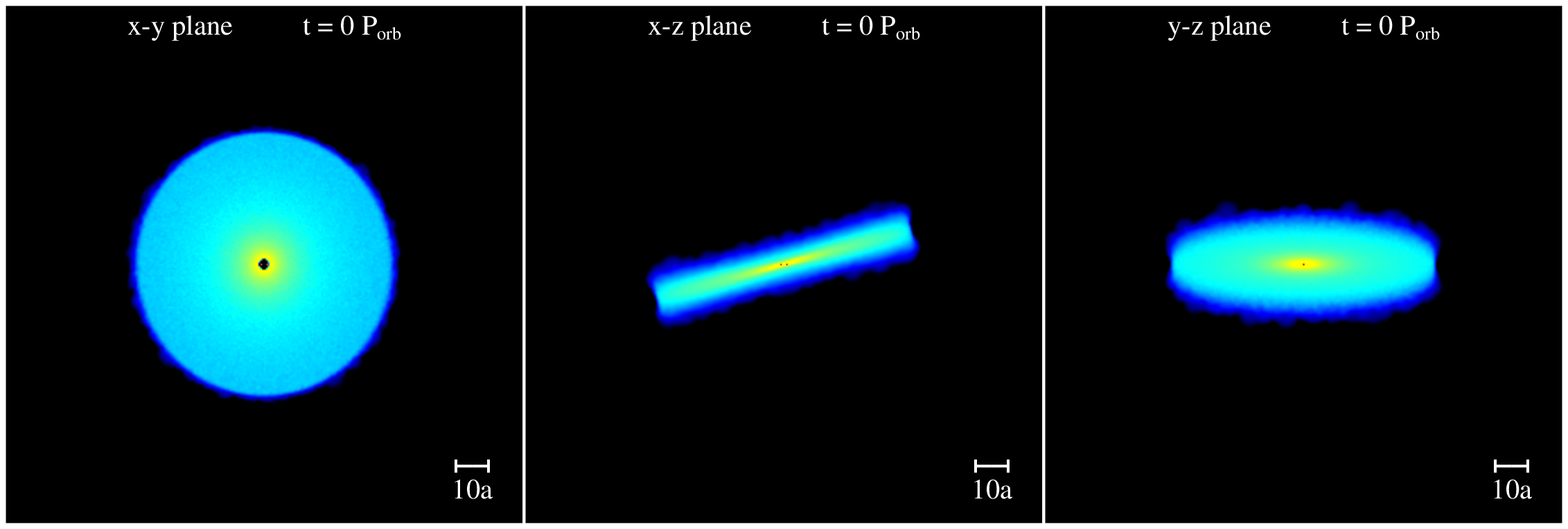}
\includegraphics[width=17.6cm]{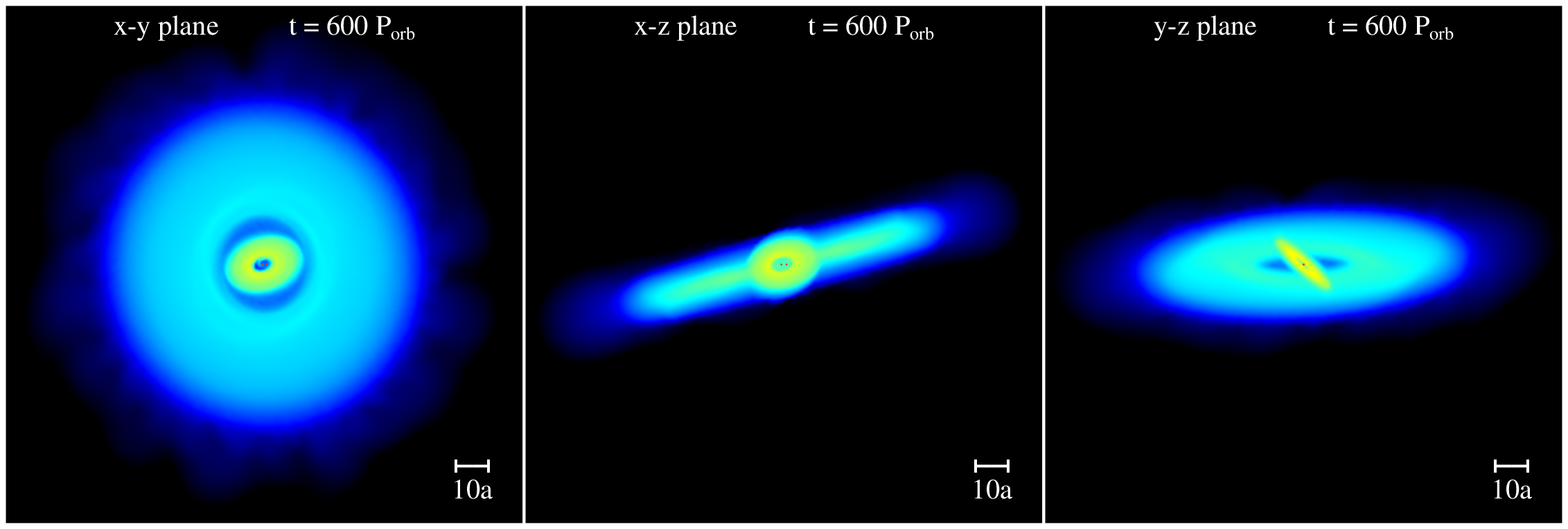}\centering
\caption{ Disc evolution for a circumbinary disk with $i_0 = 15\degree$ and $r_{\rm out}=40a$ around a binary with $e_{\rm b} = 0.8$ (Run5). Upper panels: initial setup of a low mass disc initially containing $1,000,000$ equal mass gas particles. Lower panels: the disc  at a time of $t = 600\,{\rm P_{orb}}$ The color denotes the gas density with yellow regions being about two orders of magnitude larger than the blue.  The left panels show the view looking down on to the binary orbital plane, the $x$--$y$ plane. The middle panels show the $x$--$z$ plane and the right panels show the $y$--$z$ plane.}
\label{SPH2}
\end{figure*}

\begin{figure*} \centering
\includegraphics[width=14cm]{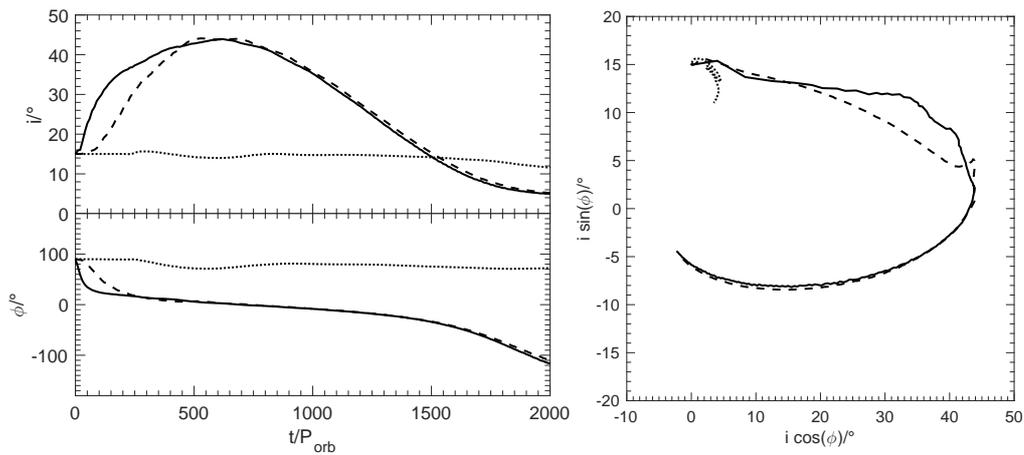}
\caption{Same as Fig.~\ref{e00} but for a circumbinary disk with $i_0 = 15\degree$ and $r_{\rm out}=40a$ around a binary with $e_{\rm b} = 0.8$ (Run5). The measurements are taken within the disc at a distance of $5a$ (solid), $10a$ (dashed), and  $25a$ (dotted).}
\label{e08_10_50a}
\end{figure*}

We consider a disc with a radial extent initially of $40a$. Unlike previous simulations in this work, this disc has $1 \times 10^6$ equal mass gas particles, more than in the other simulations, although the particle density and therefore spatial resolution is lower. The disc aspect ratio at the outer boundary is $0.047$.  Extending the disc outer radius by a factor of $8$ increases the disc angular momentum compared to the previous simulations. We investigated whether there are significant dynamical effects that the extended disc exerts on the binary. The maximum deviation from the initial binary inclination and eccentricity is $0.0072\degree$ and $0.0104$, respectively. Thus, there are no significant dynamical effects on the binary.  The initial disc setup is shown in the top panels of Fig.~\ref{SPH2}.  The evolution of the tilt and longitude of ascending node are shown in Fig.~\ref{e08_10_50a}.  
 We show the results at three radii within the disc, $5a$, $10a$ and $25a$.  For this larger disc, the sound crossing time over the radial extent of the disc is longer than the precession timescale. The inner parts of the disc begin a tilt oscillation while the outer parts of the disc remain close to their original value for longer.  The lower panels of Fig.~\ref{SPH2} show the disc at a time of $600\,P_{\rm orb}$.  The outer parts of the disc have not changed much from the initial setup, while the inner parts of the disc are significantly tilted. We see evidence for disc breaking in this simulation. %However, we do note that for higher resolution it may be possible \citep{Nixonetal2012b,NK2012,Nealon2015}.

\begin{figure} \centering
\includegraphics[width=8.4cm]{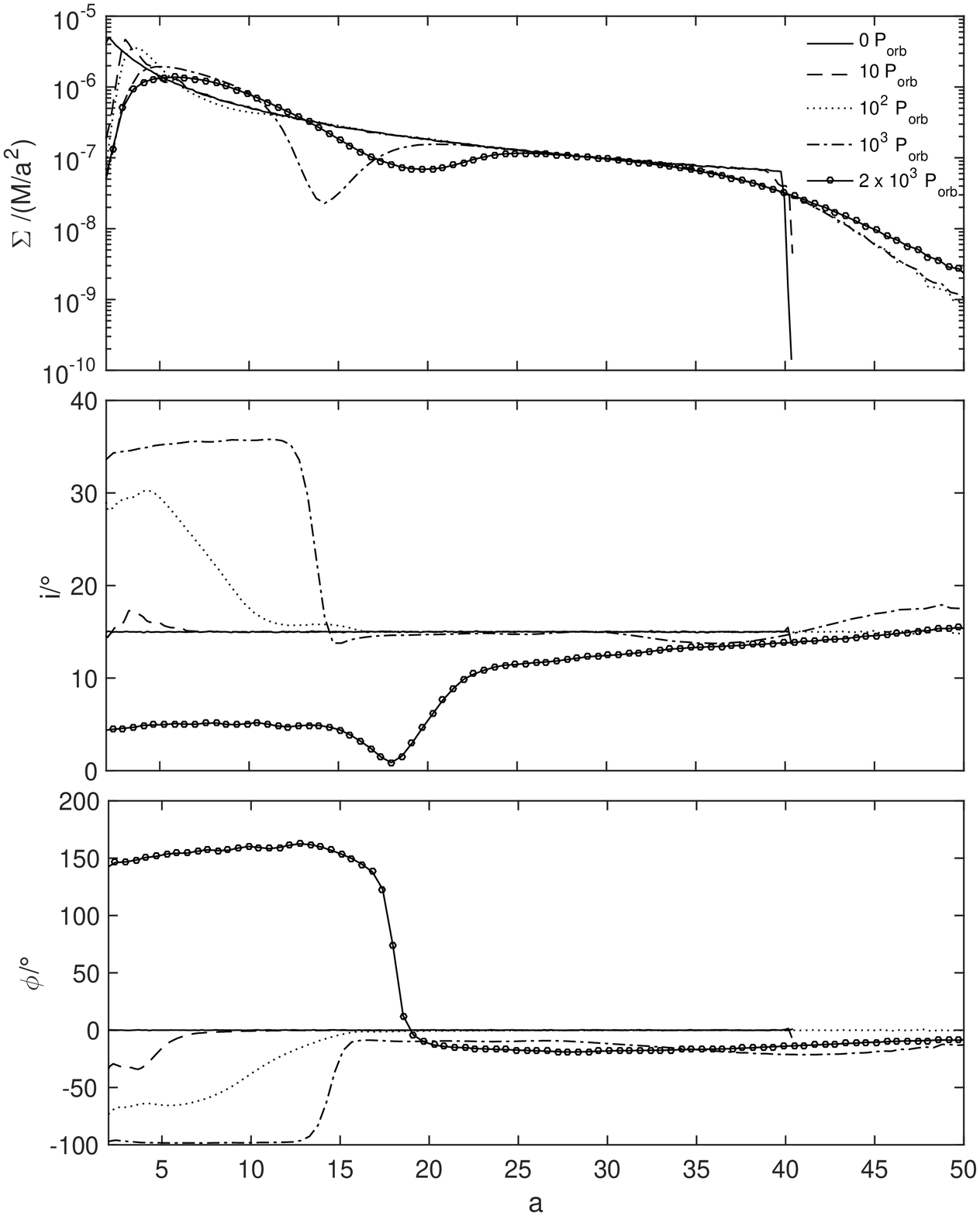}
\caption{As a function of normalized radius, we show the surface density (top panel), tilt (middle panel), and longitude of the ascending node (bottom panel) at times $0\, \rm P_{orb}$, $10\, \rm P_{orb}$, $10^2\, \rm P_{orb}$, $10^3\, \rm P_{orb}$ and $2\times 10^3\, \rm P_{orb}$. The initial conditions for the circumbinary disk are $i_0 = 15\degree$ and $r_{\rm out}=40a$ with a binary eccentricity $e_{\rm b} = 0.8$ (Run5).}
\label{sigma}
\end{figure}

To examine the behavior of the warp propagation, in Fig.~\ref{sigma} we show the surface density (top panel), inclination (middle panel), and longitude of the ascending node (bottom panel) as a function of radius at times $0\, \rm P_{orb}$, $10\, \rm P_{orb}$, $10^2\, \rm P_{orb}$, $10^3\, \rm P_{orb}$, and $2\times 10^3\, \rm P_{orb}$. The initial surface density (at $t = 0$) has a profile of $\Sigma \propto r^{-3/2}$. As the disc evolves, the gas in the outer portions of the disc spreads outwards through viscosity. As time increases, the inclination of the inner portions of the disc increases due to these tilt oscillations and the wave travels outwards in time. From the $1000\, \rm P_{orb}$ curve in the middle panel, we see that the disc below a distance of about $20a$ is inclined more than the outer regions of the disc. 
Since the surface density at  $1000\, \rm P_{orb}$ shows a dip at around $14a$, we find that the disc is  broken. 

Disc breaking occurs when the radial communication time-scale is larger than the  is the precession time-scale, $t_{\rm c} > t_{\rm p}$. The disc is able to maintain radial communication via pressure induced bending waves that propagate at speed $c_{\rm s}/2$ for gas sound speed $c_{\rm s}$  \citep{Papaloizou1995,Lubowetal2002}. The radial communication time-scale can be approximated by
\begin{equation}
t_{\rm c} \approx \frac{4}{(2+s)\Omega_{\rm b} h_{\rm out}}\bigg( \frac{r_{\rm out}}{a_{\rm b}} \bigg)^{3/2}
\end{equation}
\citep{Lubow2018}, where $h_{\rm out}$ is the disc aspect ratio at the outer edge, $s$ is related to the temperature profile of the disc ($T(r) \propto r^{-s}$), the angular frequency $\Omega_{\rm b} = \sqrt{\frac{GM}{a^3}}$.
The nodal precession rate can be approximated by
\begin{equation}
\omega_n(r) = k\bigg(\frac{a}{r}\bigg)^{7/2}\Omega_{\rm b}
\end{equation}
where
\begin{equation}
k = -\frac{3}{4}\sqrt{1+3e_{\rm b}^2-4e_{\rm b}^4}\frac{M_1 M_2}{M^2}.
\end{equation}
The precession time-scale can be found by taking the inverse of the nodal precession rate. For a narrow disc we have  $r_{\rm out} = r = 5a$, $e_{\rm b} = 0.8$, and $h_{\rm out} = 0.0795$, which equates to $t_{\rm c} \approx 160 \, \rm P_{orb} $ and $t_{\rm p} \approx 1317\, \rm P_{orb}$. Given that $t_c < t_p$, the narrow disc can rigidly precess. For example, we compare $t_{\rm p}$ to the numerical precession timescale $t_{\rm p, Run4}$ for simulation Run4 which is referenced in Fig.~\ref{e08_1}. We find that $t_{\rm p, Run4} \approx 1540\, \rm P_{orb}$ which is consistent with $t_{\rm p}$.

For a larger disc, $r_{\rm in} \ll r_{\rm out}$, the precession time-scale can be determined by taking the inverse of the global precession rate. The global precession rate of a disc is found by taking its angular momentum weighted average of the nodal precession rate $\omega_n(r)$. Therefore, the global precession time-scale is given as
\begin{equation}
t_{\rm p, global} = \frac{2(1+p)r_{\rm in}^{1+p}r_{\rm out}^{5/2-p}}{\lvert k \rvert (5-2p)a^{7/2}\Omega_{\rm b}},
\end{equation}
where $p$ is related to the initial surface density profile of the disc ($\Sigma \propto r^{-p}$),
For an extended disc with $r_{\rm out} = 40a$, $e_{\rm b} = 0.8$ and $h_{\rm out} = 0.0473$, we have $t_{\rm c} \approx 6114\, \rm P_{orb} $ and $t_{\rm p, global} \approx 2665\, \rm P_{orb} $. Since $t_{\rm c} > t_{\rm p, global}$, breaking can occur within the disc. %By equating the precession time-scale $t_{\rm p,global}$ with the sonic communication time-scale $t_{\rm c}$ we can estimate the break radius as
%\begin{equation}
%r_{\rm break} \approx \bigg( \frac{2(5-2p)\lvert k \rvert}{h_{\rm out}(2+s+2p+ps)} \bigg(\frac{a}{r_{\rm out}}\bigg)^{1-p} \bigg)^{\frac{1}{1+p}}a_{\rm b}.
%\end{equation}
%For the parameters of the extended disc, we estimate a break radius $r_{\rm break} \approx 2.79 a_{\rm b}$. This value is roughly off by a factor of 6 when compared to the break radius in the SPH simulations. This difference can be attributed to the disc being initially tilted by $15\degree$, since the above calculations assume a nearly coplanar disc. }

\section{Nearly Coplanar Disc Linear Model}
\label{sec:model}

In this section we apply a 1D linear model to the disc evolution based on equations that assume that
the level of tilt is small and that the density evolution can be ignored. The equations 
apply the secular torque due to an eccentric binary obtained by \cite{Farago2010}. The advantage of
using this approach is that solutions can be readily obtained over very long timescales for very large discs
with far less computational effort than is required with SPH. Such an approach
to modeling the circumbinary disc around KH 15D has been applied by \cite{Lodato2013} and \cite{Foucart2014}
for a circular orbit binary.
The analysis presented in this section is similar to that of \cite{Lubow2018} who analyzed  a nearly polar disc around an eccentric orbit binary.

We consider an eccentric binary with component stars that have masses $M_1$ and $M_2$ and
total mass $M=M_1+M_2$ in an orbit with semi--major axis $a_{\rm b}$ and eccentricity $e_{\rm b}$.  
To describe this configuration, we again apply a Cartesian coordinate system $(x,y,z)$ whose origin is at the binary center of mass and with the $z$-axis parallel to the binary angular momentum $\bm{J}_{\rm b}$
and the $x$-axis parallel to the binary eccentricity vector $\bm{e}_{\rm b}$. 
We consider the disc to be 
composed of circular rings that provide a  surface density $\Sigma(r)$. The ring  orientations
vary with radius $r$ and time $t$ and orbit with Keplerian angular speed $\Omega(r)$. In this model, the disc surface density 
is taken to be fixed in time, i.e., 
viscous evolution of the disc density is ignored.
We denote the unit vector parallel to the ring angular  momentum at each radius $r$ at each time $t$ by $(\ell_x(r,t),\ell_y(r,t), \ell_z(r,t))$. We consider small departures of the disc from the $x-y$ plane,
so that
$| \ell_x| \ll 1, |\ell_y| \ll 1,$ and $\ell_z  \approx 1$.
% We assume that the surface density of the disc scales as $\Sigma \propto R^{-s}$.

We apply equations~(12) and (13)
in \cite{Lubow2000} for the evolution of the disc 2D tilt vector $\bm{\ell}(r,t)=(\ell_x(r,t), \ell_y(r,t))$ and 2D internal torque $\bm{G}(r,t)=(G_x,G_y)$.  The disc tilt $i$ in radians is related to the tilt vector
by $i(r,t)= |\bm{\ell}(r,t)|= \ell(r,t)$. The tilt evolution equation is given by
\begin{equation}
\Sigma r^2 \Omega \frac{\partial \bm{\ell}}{\partial t}=\frac{1}{r}\frac{\partial \bm{G}}{\partial r}+\bm{T},
\label{lubow1}
\end{equation}
where $\bm{T}$ is the tidal torque per unit area due to the eccentric binary whose orbit lies in the $x-y$ plane.
%Note that we have dropped
%the  precessional term on the LHS of equation  (12) of \cite{Lubow2000},  since its effects are included in term $\bm{T}$.
Equation~(13) in \cite{Lubow2000} provides the internal torque evolution equation
\begin{equation}
\frac{\partial \bm{G}}{\partial t} - \omega_{\rm a} \bm{e}_z \times \bm{G}+\alpha \Omega  \bm{G}=\frac{{\cal I} r^3 \Omega^3 }{4}\frac{\partial \bm{\ell}}{\partial r},
\label{lubow2}
\end{equation}
where $\alpha$ is the usual turbulent viscosity parameter, $\omega_{\rm a}(r)$ is the apsidal precession rate for a disc that is nearly coplanar with the binary orbital plane that is given by
\begin{equation}
\omega_{\rm a}(r)= \frac{3}{8} (2+3 e_{\rm b}^2) \frac{M_1 M_2}{M^2} \left(\frac{a_{\rm b}}{r} \right)^{7/2}  \, \Omega_{\rm b} 
\end{equation}
and
\begin{equation}
{\cal I} = \int \rho z^2 dz,
\end{equation}
for disc density $\rho(r)$.
We apply boundary conditions that the internal torque vanishes at the inner and outer disc edges $r_{\rm in}$ and $r_{\rm out}$,
respectively. That is,
\begin{equation}
\bm{G}(r_{\rm in}, t) = \bm{G}(r_{\rm out}, t) =0.
\label{GBC}
\end{equation} 
 This is a natural boundary condition because the internal torque vanishes just outside the disc boundaries. Thus, any smoothly
varying internal torque would need to satisfy this condition.

The torque term due to the eccentric binary  follows from an application of
  equations (2.17) and (2.18) in \cite{Farago2010}.  The torque term is
expressed as 
\begin{equation}
\bm{T}=\Sigma r^2 \Omega \bm{\tau}
\label{Ttor}
\end{equation}
with 
\begin{equation}
\bm{\tau}=(a(r) \ell_y, b(r) \ell_x) 
\label{tau}
\end{equation}
and
\begin{equation}
a(r) = (1 -e_{\rm b}^2) \, \omega_{\rm p}(r), 
\label{taux}
\end{equation}
and
\begin{equation}
b(r)=-(1+4e_{\rm b}^2) \, \omega_{\rm p}(r),
\label{tauy}
\end{equation}
where frequency $\omega_{\rm p}$ is given by
\begin{equation}
\omega_{\rm p}(r)=\frac{3}{4}\frac{M_1 M_2}{M^2} \left(\frac{a_{\rm b}}{r} \right)^{7/2} \Omega_{\rm b}.
\label{omegap}
\end{equation}

We seek solutions of the form $\bm{\ell} \propto e^{i\omega t}$ and $\bm{G} \propto e^{i\omega t}$ and Equations (\ref{lubow1}) and (\ref{lubow2}) become
\begin{equation}
 i  \omega   \Sigma r^2 \Omega \bm{\ell} =\frac{1}{r}\frac{d \bm{G}}{d r}+  \Sigma r^2 \Omega\bm{\tau}
\label{l1}
\end{equation}
and
\begin{equation}
i \omega \bm{G} - \omega_{\rm a} \bm{e}_z \times \bm{G}+\alpha \Omega  \bm{G}=\frac{{\cal I} r^3 \Omega^3}{4} \frac{d \bm{\ell}}{d r},
\label{l2}
\end{equation}
respectively. As usual, the physical values of $\bm{\ell}$ and $\bm{G}$ are obtained by taking their real parts.

\section{Nearly Rigid Disc Expansion}
\label{sec:expansion}
\subsection{Lowest order}

We apply the nearly rigid tilted disc expansion procedure in \cite{Lubow2000}.
We  expand variables in the tidal potential that is considered to be weak as follows:
\begin{eqnarray}
a &=& A^{(1)},\\
b &=& B^{(1)}, \nonumber \\
\bm{\ell} &=& \bm{\ell}^{(0)} + \bm{\ell}^{(1)} + \cdots,  \nonumber \\
\omega &=& \omega^{(1)} + \omega^{(2)} + \cdots,  \nonumber \\
\bm{G} &=& \bm{G}^{(1)} + \bm{G}^{(2)} + \cdots,  \nonumber \\
\bm{\tau} &=& \bm{\tau}^{(1)} + \bm{\tau}^{(2)} + \cdots,   \nonumber
\end{eqnarray}
where $a$ and $b$ are given by Equations (\ref{taux}) and  (\ref{tauy}), respectively.  $a$ and $b$ depend on the tidal
potential and are regarded as first order quantities.

To lowest order, the disc is rigid and the tilt vector $\bm{\ell}^{(0)} = (\ell_{x}^{(0)}, \ell_{y}^{(0)})$ is constant in radius.
We integrate $r$ times Equation (\ref{l1}) over the entire disc and apply the boundary conditions given by Equation (\ref{GBC}) to obtain
\begin{equation}
\int_{r_{\rm in}}^{r_{\rm out}} \Sigma r^3 \Omega(i\omega^{(1)} \bm{\ell}^{(0)}- \bm{\tau}^{(1)})\, dr=0,
\end{equation}
where  
\begin{equation}
\bm{\tau}^{(1)}(r) = (A^{(1)}(r) \, \ell_{y}^{(0)}, B^{(1)}(r)\, \ell_{x}^{(0)}).
\end{equation}
We then obtain for the disc precession rate in lowest order 
\begin{equation}
\omega^{(1)} = -\frac{3}{4}  \sqrt{1+3 e_{\rm b}^2-4 e_{\rm b}^4} \frac{M_1 M_2}{M^2}\left<\left(\frac{a_{\rm b}}{r} \right)^{7/2} \right>  \Omega_b,
\label{om1}
\end{equation}
where the bracketed term involves the angular momentum weighted average
\begin{equation}
 \left<\left(\frac{a_{\rm b}}{r} \right)^{7/2} \right>= \frac{ \int_{r_{\rm in}}^{r_{\rm out}}\Sigma r^3 \Omega  (a_{\rm b}/r)^{7/2} dr}{ \int_{r_{\rm in}}^{r_{\rm out}} \Sigma r^3 \Omega dr}.
 \label{javg}
\end{equation}
We define the precession period as
\begin{equation}
P_{\rm p} = \frac{2 \pi}{|\omega^{(1)}|}.
\end{equation}

The tilt components are related by
\begin{equation}
\ell^{(0)}_y= - i \sqrt{\frac{1- e_{\rm b}^2}{1+4e_{\rm b}^4}}\,  \ell^{(0)}_x.
\label{ly0}
\end{equation}
Because $|\ell^{(0)}_x|$ and $|\ell^{(0)}_y|$ differ, the disc undergoes nonuniform precession and secular tilt oscillations with tilt
variations $i(t)$ with respect to the $x-y$ plane.
The disc longitude of ascending node $\phi$
is related to the tilt vector by
\begin{equation}
\tan{(\phi(t))} = -\frac{Re(\ell^{(0)}_{\rm x}(t))}{Re(\ell^{(0)}_{\rm y}(t))}.
\end{equation}

\begin{figure}
\centering
\includegraphics[width=8.0cm]{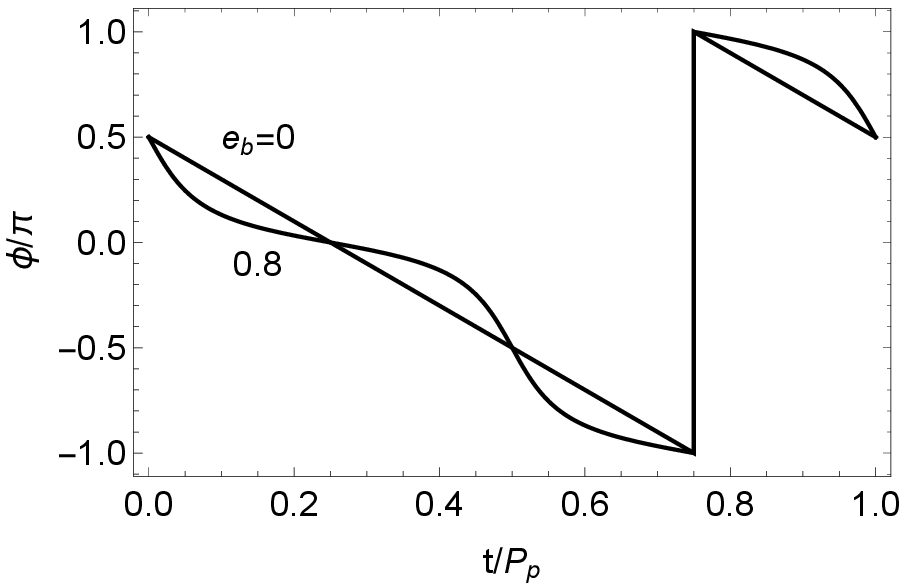}
\includegraphics[width=8.0cm]{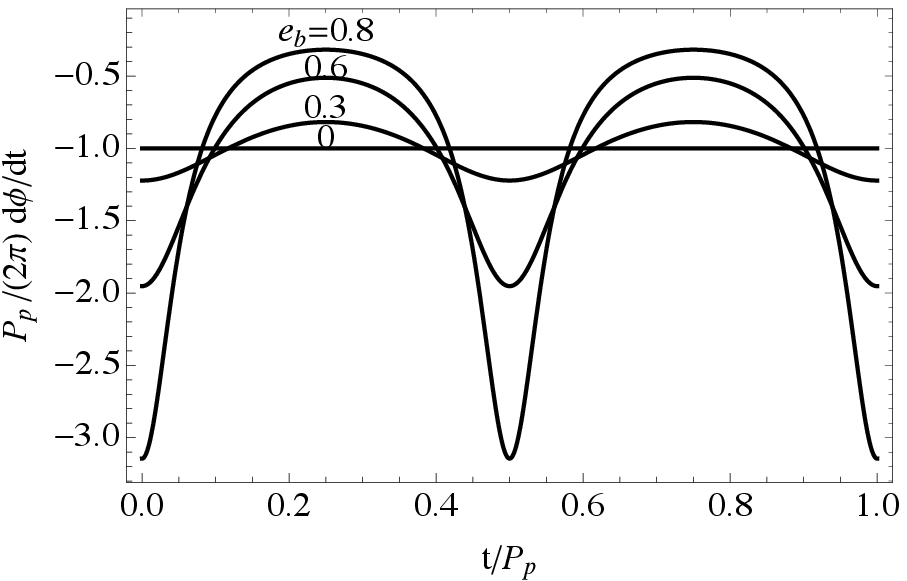}
\caption{Top panel: Longitude of ascending node $\phi$ in radians as a function of time for different values
of binary eccentricity. Bottom: Normalized nodal precession rate  as a function of time for 
various values of binary eccentricity.  }
\label{precrate}
\end{figure}

We take the initial disc longitude of ascending nodes to be  $90^\degree$, so that 2D tilt vector $\bm{\ell}$ is initially aligned with the  binary eccentricity vector.
Figure~\ref{precrate} plots the longitude of the ascending node and the nodal precession rate as a function of time for various
values of binary eccentricity. For $e_{\rm b}=0$, the precession rate is uniform and appears as the horizontal
line. The precession rate becomes highly nonuniform at higher values of binary eccentricity. 
For $e_{\rm b}=0.8$, the precession rate varies about a factor of 10 over the precession period.

The results in the upper panel of Figure~\ref{precrate} for $e_{\rm b} =0.8$ are similar to those in the lower left panel of Figure~\ref{e08_1} that are based on SPH simulations. The precession is nonuniform in both cases, with similar phase oscillations in time.
One difference is that the precession period increases
in time in the SPH simulations. This increase occurs because of the viscous disc density evolution that 
in turn changes the disc angular momentum. This effect is not included in the linear model.

\begin{figure}
\centering
\includegraphics[width=8.0cm]{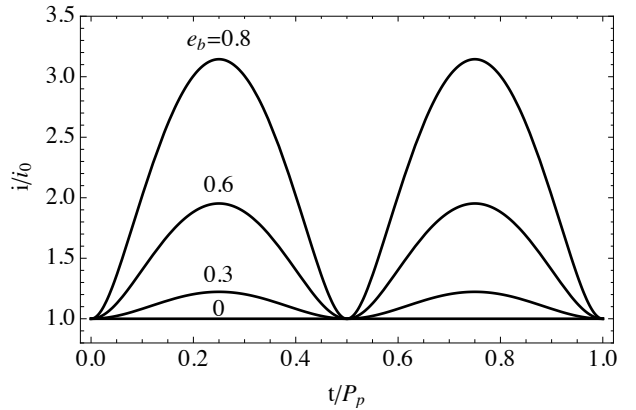}
\caption{Normalized disc tilt angle in radians relative to the coplanar orientation (Equation (\ref{it})) as a function of time for various values of binary eccentricity.  Time $t=0$ corresponds to the disc longitude of ascending node
$\phi=90^\degree$. }
\label{ivrst}
\end{figure}

The disc tilt varies in time as
\begin{equation}
i(t) = i_0 \sqrt{ \frac{ 2 + (3- 5 \cos { ( 2 \omega^{(1)} t )})e_{\rm b}^2 }{2(1-e_{\rm b}^2)}}.
\label{it}
\end{equation}
where $i_0=i(0)$ that occurs when the longitude of the ascending node is $90^\degree$.
Figure~\ref{ivrst} plots the tilt angle as a function of time for 
various values of binary eccentricity.
Tilt oscillations occur because the binary potential is nonaxisymmetric around the direction of the binary angular momentum  vector
(the $z-$axis).
For $e_{\rm b} \simeq 1$, the oscillations undergo extreme tilt variations $i(t) \propto i(0) \sqrt{(1- \cos{(2 \omega^{(1)} t)})/(1-e_{\rm b})}$.

The normalised disc tilt and precession rates plotted in Figures~\ref{precrate} and \ref{ivrst} are independent
of the disc properties such as its density and temperature distributions, provided that
the level of disc warping is small, i.e., $\bm{\ell}(r, t)$ is nearly constant in radius.
%The results in Figure~\ref{ivrst} for $e_{\rm b} =0.8$ are similar to those in the upper left panel of Figure~\ref{e08_1} that are based on SPH simulations. The tilt increases to a maximum values that is about a factor of 3 larger than the initial value in both cases.
 
\begin{figure}
\centering
\includegraphics[width=8.0cm]{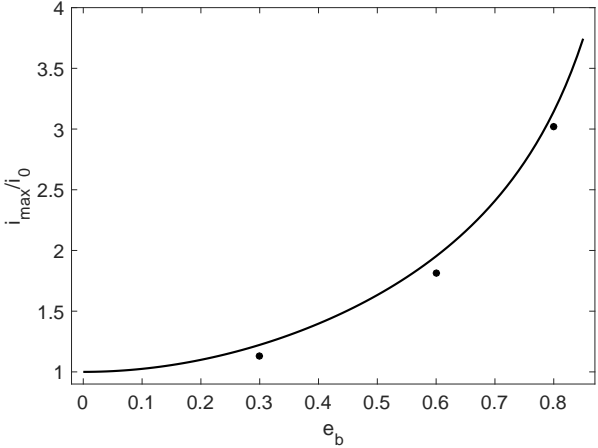}
\caption{
Plotted as a  line is the maximum disc tilt as a function of binary eccentricity 
normalized by $i_0$ (defined in Figure \ref{ivrst}) based on Equation (\ref{imax}). Plotted as dots are the results of SPH simulations for models Run6, Run7, and Run8 given in Table~\ref{table2}, which all have an initial circumbinary disk tilt $i_0 = 10\degree$.}
\label{fig:imax} 
\end{figure}

Figure \ref{fig:imax} plots the maximum
tilt angle over time as a function of binary eccentricity implied by Equation (\ref{it})
that occurs for $\cos { ( 2 \omega^{(1)} t )})=-1$,
\begin{equation}
i_{\rm max}= i_0\sqrt{ \frac{ 1 + 4 e_{\rm b}^2 }{1-e_{\rm b}^2}}.
\label{imax}
\end{equation}
Also plotted on the figure are the maximum inclinations for SPH simulations for models listed in Table 1
that start with $i_0=10^ \degree$. As seen in the figure, the results of the SPH
simulations agree well with the expected results based on linear theory. The plotted SPH  results lie slightly below the
expectations of linear theory, likely due to the effects of disc dissipation.  Though the linear theory is valid for low inclinations, the SPH simulations that begin at higher inclinations, $i_0 \le 50\degree$ (Runs 2, 3, and 4) also agree quite well with the linear model.

\section{Circumbinary disc of KH 15D}
\label{kh15d}

KH 15D is a spectroscopic binary T Tauri star in the cluster NGC 2264 and located at a distance of 760 pc \citep{Sung1997}. This system was originally thought to be a single variable star. But more in-depth observations showed this system had a stellar companion, which causes a peculiar light curve \citep{Kearns1998}. The system is estimated to be an age of  $3\times 10^6\, \rm yr$ and the total mass is roughly $1.3\, M_{\odot}$ \citep{Hamilton2001,Johnson2004}. The spectral classification of star A is K6/K7 \citep{Hamilton2001} and star B is K1 \citep{Capelo2012}. The two stellar companions are of roughly equal mass and are on highly eccentric orbits embedded in the accretion disc which emits bipolar outflows \citep{Hamilton2003,Deming2004,Tokunaga2004,Mundt2010}.  The binary has an eccentricity in the range of $e_{\rm b} =0.68$ to $0.8$ \citep{Johnson2004}, semi--major axis of $0.26\,\rm au$.

The light curve of KH 15D undergoes periodic eclipses in which the brightness drops by about $3.5$ magnitudes for a duration of roughly $24$ days with a orbital period of $48.37\, \rm days$ \citep{Johnson2004,Winn2004,Hamilton2005}. The duration of the eclipse has varied over time (see Fig.1 in \cite{Aronow2018} for the $I$-band light curve of KH 15D which shows the brightness of the system from $1951$ to $2017$). The brightness increased between 1995 and 2005 and the peak brightness decreased between 2006 and 2010 \citep{Hamilton2001,Hamilton2005}.  

To understand what causes this light curve, \cite{Chiang2004} and \cite{Winn2004} independently developed a model in which a circumbinary disc  or ring that is misaligned to the orbital plane of an eccentric orbit blocks light from the binary and undergoes nodal precession. % as a rigid body due to the tidal influence of the binary on the misaligned disc \citep{papaloizouterquem1995,Larwood1997, Lubow2000}. 
The nodal precession explains the time variations of the observed light curves.  Between 1995 and 2010, the leading edge of the disc precessed across the orbit of Star A, while star B was fully occulted. During the time between 2010 and 2012, both stars A and B were only detectable through scattered light. Currently, the brightness of the system has increased as star B's orbit has become uncovered from the the trailing edge of the precessing disc \citep{Capelo2012,Windemuth2014,Arulanantham2016,Aronow2018}.

  Previous hydrodynamical models for a gaseous disc in KH 15D have only modeled the binary as circular \citep{Lodato2013, Foucart2014}.
Our goal in modeling this system is to understand the properties of the disc, such as its radial extent, given the observed constraints. 
Based on the work by \cite{Chiang2004} and \cite{Winn2004} we consider the disc to be observed nearly edge-on
and inclined relative to the orbit of the binary. In addition, the binary eccentricity vector lies in the plane of the sky. %We assume that the orbital plane of the binary is not close to being in the plane
%of the sky.
Under these conditions, the line of ascending nodes of the disc should currently be $\phi \approx 90^\degree$.

We consider a model in which the disc tilt $i$ 
is below the critical value $i_{\rm crit}$ given by Equation (\ref{icrit}) which implies that $20^\degree \la i_{\rm crit} \la 30^\degree$ for
$0.6 \la e_{\rm b} \la 0.8$ \citep{Johnson2004}.   If the disc tilt is above this critical value, then the disc will evolve to a polar (perpendicular) alignment with the binary \citep{Martin2017}. However, for this work, we only examine the conventional model where the disc or ring is precessing about the binary angular momentum vector. 
%The polar alignment timescale tends to be much shorter than the timescale for co-planar alignment. Given the age of the system, we focus on parameters that drive the disc to align with the binary orbital plane.

We see from the Figures \ref{e08_1} and \ref{precrate} that the precession rate is largest in magnitude
at this phase $\phi \simeq 90^\degree$. For a binary eccentricity of $e_{\rm b} =0.8$, the precession rate is about $3$ times faster than the mean
precession rate. The tilt at this phase is at a minimum value. At later times the retrograde precession rate $- d\phi/ dt$ will
be as much as an order of magnitude smaller and the tilt will be more than 3 times larger. These results are largely independent of the details of the disc/ring structure. 

The observed occultation involves scattering by solid particles. Such particles
would undergo differential precession of the orbits in the presence of the binary
that would destroy the disc structure over time. Some mechanism
is required to maintain the disc flatness. One possibility is the ring coherence is maintained
by self-gravity in analogy to planetary rings \citep{Chiang2004}. Another possibility is that
the solids are coupled to a gas disc that maintains its flatness by pressure effects \citep{papaloizouterquem1995,Larwood1997, Lubow2000}. We analyze the latter model.

To analyze the system further, we numerically solve Equations (\ref{lubow1}) and (\ref{lubow2}) subject to boundary conditions given in Equation (\ref{GBC})
for disc modes, as is described in \cite{Lubow2018}.
We analyze discs whose parameters are listed in Table 1,
where $s$ and $p$ are defined by $T(r) \propto r^{-s}$ and $\Sigma(r) \propto r^{-p}$, respectively.
In all cases we assume an equal mass binary $M_1=M_2$. 
%We take $r_{\rm in} = 4 a$ for the highly eccentric binary as approximately indicated by \cite{Artymowicz1994}. 
The disc inner radii should increase somewhat with binary eccentricity, but we ignore that effect
for the two values of eccentricity being considered.

\begin{table}
 \caption{Model Parameters}
 \label{tab:ModelParams}
 \begin{tabular}{lccccccc}
  \hline
  Model & $r_{\rm in}/a$ & $H/r(r_{\rm in})$ & $\alpha$ & $p$ & $s$ &    $e_{\rm b}$ \\
  \hline
  A & 4 &0.1 &  0.01   & 0.5  & 1.0   & 0.6 \\
    \hline
    B & 4 & 0.1  &  0.01  & 0.5  & 1.0   & 0.8 \\
    \hline
      C & 4 &0.1 &  0.01    & 1.0  & 1.0   & 0.6 \\
    \hline
    D & 4 & 0.1 &  0.01   & 1.0  & 1.0  & 0.8 \\
        \hline
    E & various & 0.1 &  0.01   & 0.5  & 1.0  & 0.75 \\
            \hline
    F & various & 0.1 &  0.01   & 1.0  & 1.0  & 0.75 \\
 \end{tabular}
\end{table}

\subsection{Precession period constrained model}
\label{sec:pconst}

Previous disc models for this system by \cite{Lodato2013} and \cite{Foucart2014} applied a constraint on the disc precession period based on the \cite{Chiang2004} model. In that model, the precession period is approximately $3000\,\rm yr$
or about $2.09 \times 10^4 P_{\rm b}$. However, this period value is determined by considering a narrow ring
 and so it is not clear how well this constraint would apply to a broad disc.  
This model may be appropriate if the occultation
is due to material in the somewhere in the middle  of the radial extent of disc, rather than
the outer edge.
 We consider an alternate model in the next subsection. We describe results for a disc period constrained model based on results from linear modes.

\begin{table}
 \caption{Period Constrained Results}
 \label{tab:Pconstrained}
 \begin{tabular}{lccc}
  \hline
  Model & $r_{\rm out}/a$ & $\tau$ (yr) & Max($r/\ell_{\rm in} |d \ell/dr|$) \\
  \hline
  A & 27.6 & $2.3 \times 10^5$ & 0.04 \\
    \hline
    B & 26.4 & $2.8 \times 10^5$  & 0.06    \\
    \hline
      C & 37.3 & $1.3 \times 10^5$   & 0.05 \\
    \hline
    D & 35.0 & $1.6 \times 10^5$ & 0.07   \\
 \end{tabular}
\end{table}

We adopt the disc parameters similar to those of \cite{Lodato2013} that are listed for Models A-D in Table \ref{tab:ModelParams}.
In addition we  consider two values of binary eccentricity $e_{\rm b} =0.6$ and 0.8, while
the previous models considered a circular orbit binary.
Table~\ref{tab:Pconstrained} contains results
for these models.
The columns in the table are for the values for the disc outer radius $r_{\rm out}/a$, decay timescale of the tilt
in year $\tau$,
and the maximum normalized warp value across the disc Max($r/\ell_{\rm in} |\ell/dr|$). The latter is the magnitude of the
logarithmic radial derivative of the tilt vector $\bm{\ell}$ divided by the magnitude of the tilt
at the disc inner edge, $\ell_{\rm in}$ (see also Section 3.2 of \cite{Martin2018} for more details).
Since this value is small, less than $H/r$, for all disc models, the disc warp is very mild and so
the disc behaves quite rigidly. In addition, the linear treatment of the disc evolution is well justified
for discs with small tilts.
 
The numerical results are similar to those in \cite{Lodato2013} and \cite{Foucart2014} once slight differences
in the model parameters are taken into account. For example, Table 1 in \cite{Lodato2013} has a value
for $r_{\rm out} = 26 a$ for $p=0.5$, while we obtain a value of 27.6 in Model A. The small difference is likely due to binary eccentricity and
the slightly different value of the binary semi-major axis adopted. In any case, as obtained previously, the disc model decays rapidly
compared to the system lifetime of a few million years. The decay rate is proportional to the $\alpha$ value in the disc (for a fixed disc structure) and suggests
that reductions to $\alpha \sim 10^{-3}$ are required to provide a sufficiently slow tilt decay.

The effect of binary eccentricity is to slightly decrease the required disc outer radius, as seen in comparing
Models A and B and also Models C and D. In addition the decay timescale slightly increases with increasing
binary eccentricity. 

\subsection{Velocity constrained model}
\label{sec:vconst}

There is an observational constraint 
on the speed of
the occulting disc/ring in the plane of the sky.
By comparing frames 1
and 4 in Figure 1 of \cite{Aronow2018}, we estimate that the occultation occurs across
distance $\simeq a (1+e_{\rm b})$ over a time $\tau_0$ of roughly 40 years. If we take the standard value of $a=0.26$ AU, we then have a constraint on the transverse occulting velocity $v \sim a (1+e)/\tau_0$, that is   
\begin{equation}
v \simeq 6.5 \times 10^{-3} (1+e_{\rm b}) \, { \rm AU/yr}. 
\label{vobs}
\end{equation}
As discussed above, this velocity occurs for the longitude of ascending nodes that we take $\phi=90^\degree$. We apply this velocity constraint for various models computed from linear modes.

\begin{figure}
\centering
\includegraphics[width=8.0cm]{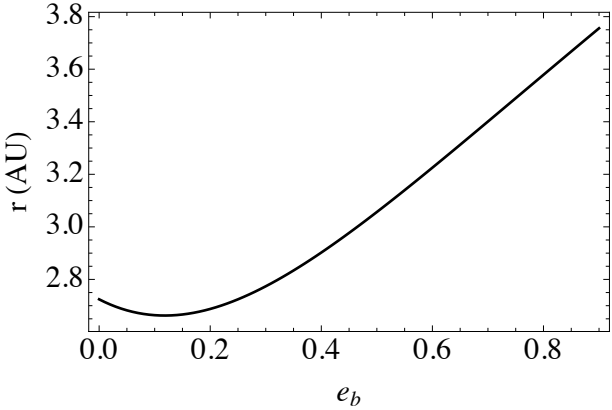}
\caption{Narrow ring radii that satisfy the velocity constraint described in Section \ref{sec:vconst} as a function of binary eccentricity. }
\label{fig:er}
\end{figure}

For a narrow ring, we determine
the ring radii as a function of binary eccentricity that satisfy the velocity constraint (\ref{vobs}) at $\phi=90^\degree$.
The results are plotted in Figure~\ref{fig:er}.
The radii agree well with the $\sim 3$ AU estimated by \cite{Chiang2004}. For larger values of binary eccentricity, the ring radius increases with eccentricity.

\begin{figure}
\centering
\includegraphics[width=8.0cm]{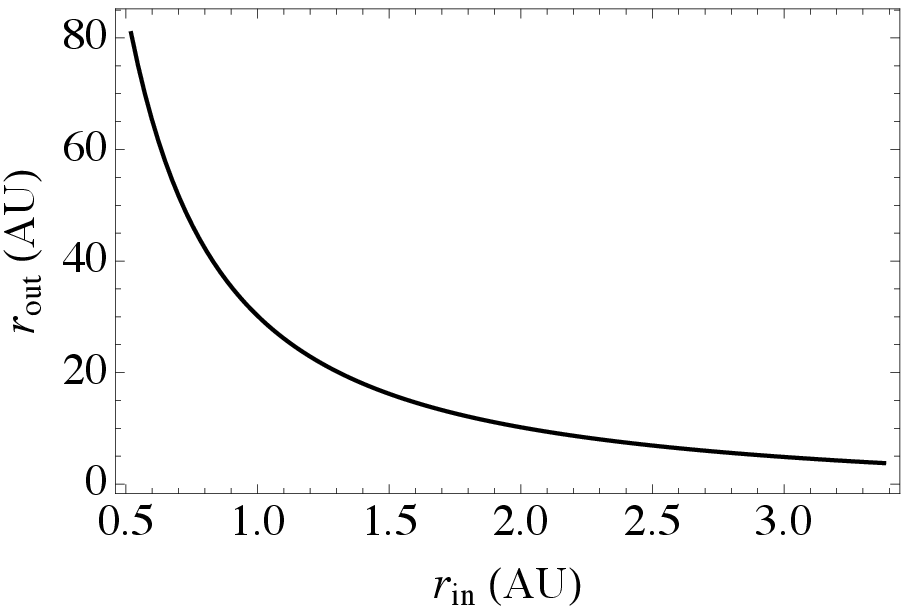}
\caption{Disc outer radius as a function of disc inner radius for
a disc with $p=0.5$ and binary eccentricity $e_{\rm b} =0.75$ that satisfies the velocity constraint described in Section \ref{sec:vconst}. }
\label{fig:rinroutp0p5}
\end{figure}

For a broader disc, we assume the occultation is dominated by the disc outer edge. 
We then apply the velocity constraint at
that radius. In Figure~\ref{fig:rinroutp0p5}
we plot the disc outer radius as a function
of disc inner radius 
for Model E of Table \ref{tab:ModelParams}
that has a disc
with surface density parameter $p=0.5$ and assumed binary eccentricity $e_{\rm b} =0.75$. The value of $e_{\rm b}$ is close to the best fit value of 0.74 in the model of \cite{Johnson2004}.

The inner radius of the circumbinary 
disc in KH 15D is expected to range roughly
from $r_{\rm}= 0.5 {\rm AU}$ at higher
viscosities $\alpha > 0.01$ to $r_{\rm} = 1 {\rm AU}$ at small viscosities $\alpha < 1\times 10^{-5}$
due to the balance of viscous torque with tidal torques 
\citep{Artymowicz1994}.
The disc torque increases for smaller disc inner radii and  is insensitive to the disc outer radius for $r_{\rm in} \ll r_{\rm out}$.  The disc angular momentum increases with the disc outer radius.
For smaller disc inner radii, there is a stronger torque due to the binary that requires a larger disc outer radius to produce the same velocity at the disc outer edge. There is then an inverse relationship between the
inner and outer disk radii. 

\begin{figure}
\centering
\includegraphics[width=8.0cm]{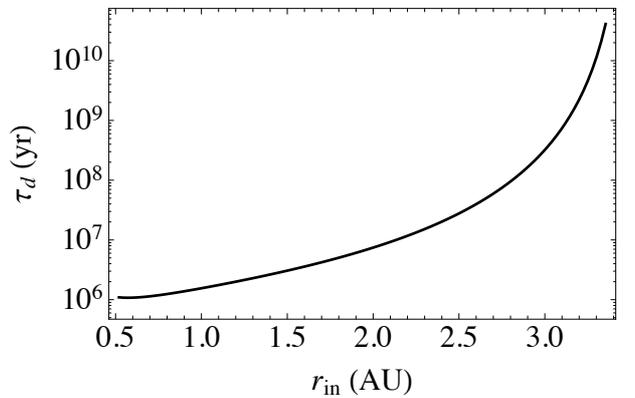}
\caption{Tilt decay time $1/Im(\omega)$ as a function of disc inner radius for
a disc with $p=0.5$ and binary eccentricity $e_{\rm b} =0.75$ that satisfies the velocity constraint described in Section \ref{sec:vconst}. }
\label{fig:tdp0p5}
\end{figure}

In Figure~\ref{fig:tdp0p5}, we plot the tilt decay timescale for Model E of Table \ref{tab:ModelParams}
as a function of disc inner radius with parameters $s=1.0, \alpha=0.01$, and $H/r(r_{\rm in})=0.1$.  In this case, the disc 
decay timescale is typically of order the disc lifetime of a few million years or longer. The velocity constrained model undergoes slower tilt decay than the similar models for the period constrained models of Section \ref{sec:pconst}. In particular, no reduction of $\alpha$ below $0.01$ is required in this case to meet the requirement that the disc decay timescale exceed the disc lifetime.

\begin{figure}
\centering
\includegraphics[width=8.0cm]{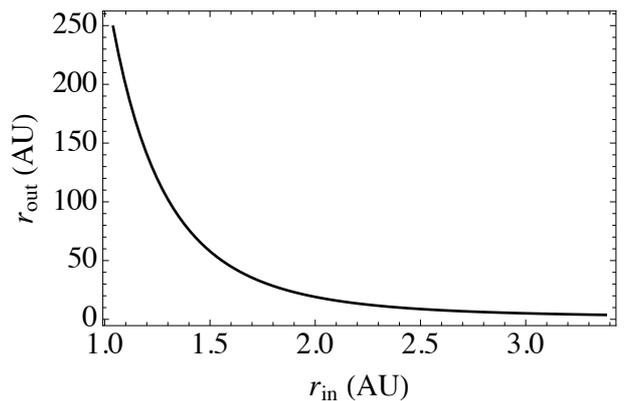}
\caption{Disc outer radius as a function of disc inner radius for
a disc with $p=1$ and binary eccentricity $e_{\rm b} =0.75$ that satisfies the velocity constraint described in Section \ref{sec:vconst}. }
\label{fig:rinroutp1p0}
\end{figure}

We now consider the velocity constrained model with Model F in Table \ref{tab:ModelParams} that has the same parameters as Model E, but with $p=1$.
In this case, the disc outer radius is required to be considerably larger than the $p=0.5$ case, as seen in Figure~\ref{fig:rinroutp1p0}.
We limited the plot to $r_{\rm in} \ge 1 {\rm AU}$ because at smaller values of $r_{\rm in}$ the disc outer radius gets very large.
The reason is that the surface
density falls off faster with radius. The increased radius in the $p=1$ case is required to produce a large enough disc angular momentum that is sufficient to reduce the disc velocity at the outer edge in order the meet the velocity constraint.
We find that the tilt decay timescale with $p=1$ is even longer than indicated
in Figure~\ref{fig:tdp0p5}. Again, no reduction in adopted $\alpha=0.01$ is required for the tilt to survive a few million years.

These models have assumed that the occultation occurs due to
material at the gaseous disc outer edge. The occultation is likely due to solids (dust)
that could have migrated inward somewhat from the gaseous disc outer edge. This effect would make the velocity constraint easier to satisfy. That is, the gas disc outer radius could be smaller
than indicated in Figures~\ref{fig:rinroutp0p5} and \ref{fig:rinroutp1p0} and satisfy the velocity constraint 
of Equation (\ref{vobs}). The level of reduction for $r_{\rm out}$ depends
on the degree to which the solids have migrated inward, as is discussed in Section \ref{sec:thick}.

\subsection{Constraint on thickness of obscuring layer}
\label{sec:thick}

 The obscuring material
likely consists of solids that form a dust embedded layer
within the gaseous disc. IR observations 
suggest that the solids consists of 1 to 50 micron size particles
\citep{Arulanantham2016, Arulanantham2017}. 
We define the full thickness of the 
obscuring layer as $2 T$.
The observations of KH 15D show that both stars were occulted over a time interval $\tau \sim 5$ years  \cite[see Figure ~1 of][]{Aronow2018}.
The disc thickness can then be expressed as 
\begin{equation}
2 T \simeq a (1+e_{\rm b}) (1+ \tau/\tau_0) \sin{i},
\end{equation}
where the term involving $\tau$ is due to the transverse velocity (precession) of the disc given in Equation (\ref{vobs})
and $\tau_0$ is the time for the disc leading edge to precess across both stars that we estimate as $\tau_0 \sim 40$ years, as discussed in Section \ref{sec:vconst}.
The term involving $\tau$ is then a small correction $\sim 10\%$ that we ignore. The constraint on $T$
then implies that
\begin{equation}
T \sim 0.13 (1+e_{\rm b}) \sin{i} \, {\rm AU}. 
\end{equation}

We consider how this constraint applies
to the velocity constrained model of Section \ref{sec:vconst}.
For the narrow ring case with $e_{\rm b}=0.75$ and $r_{\rm out}= 3$ AU
(see Figure~\ref{fig:er}), we have then $T/r_{\rm out} \sim 0.1 \sin{i}$.
For $e_{\rm b}=0.75$ and a circumbinary disc with $p=0.5$ that is tidally truncated by the binary at its inner radius at $r_{\rm in} \sim 1$ AU,
we have from Figure~\ref{fig:rinroutp0p5} that $r_{\rm out} \sim 30$ AU and so $T/r_{\rm out} \sim 0.01 \sin{i}$.
For a circumbinary disc with the same set of parameters, but with $p=1$,
we have that % $r_{\rm out} \sim xx$ (see Figure~\ref{fig:rinroutp1p0})and so 
$T/r_{\rm out} \sim 0.001 \sin{i}$.
For a  narrow ring, the thickness of the occulting solid layer is comparable to the thickness
of the gaseous disc layer, if $\sin(i)$ is not small, which suggests that mild settling of solids has occurred. 
But, the broad disc $T/r_{\rm out}$ values are significantly smaller than the assumed gas disc
aspect ratio $H/r \sim 0.1$, typical of protostellar disc aspect ratios. Such small $T/r_{\rm out}$ values suggest that settling of solids towards the disc midplane has occurred. Such settling suggests that the radial
drift of solids might have also occurred so that the velocity constraint  may be satisfied
with a smaller gaseous disc outer radius, as discussed in Section \ref{sec:vconst}. 

To produce such thin layers in the broad disc cases of Section \ref{sec:vconst} requires
that the level of disc turbulence be very low.
Using equations 19 and 20 of \cite{Fromang2009} and setting the Schmidt number to unity, we estimate
that
\begin{equation}
\alpha \sim \Omega \, t_{\rm s} \left(\frac{T}{H}\right)^2,
\label{als}
\end{equation}
where $t_{\rm s}$ is the stopping time for the particles
given by equation 10 of \cite{Fromang2009}.
For $\sim 50$ micron particles and $H = 0.1 r$,
we have that the upper limit to $\alpha$
is 
\begin{equation}
\alpha \sim \alpha_0   \left( \frac{r_{\rm out}}{r} \right)^{2-p} \left( \frac{0.001 M_\odot}{M_{\rm d}} \right) (1+e_{\rm b})^2 \sin^2{i}
\label{allim}
\end{equation}
where 
\begin{equation}
\alpha_0 = \frac{1.5 \times 10^{-5}}{2-p}.
\end{equation}
\cite{Aronow2018} report an upper limit of the disc mass as $\simeq 1.7 \times 10^{-3} M_\odot$ based on ALMA nondetections.
For the outer parts of the velocity constrained disc in Figure \ref{fig:rinroutp0p5} with $p=0.5$ and $r_{\rm in} =1$ AU and $r=r_{\rm out}=30$ AU, 
 we obtain for a disc with $M_{\rm d}= 0.001 M_\odot$ and $e_{\rm b}=0.75$ from  Equation ({\ref{allim}}) that
$\alpha \sim 10^{-5} \sin^2{i}$. 
For outer parts of the velocity constrained disc in Figure \ref{fig:rinroutp1p0} with $p=1.0$ and $r_{\rm in} =1$ AU and $r=r_{\rm out}=170$ AU,  we obtain for a disc with $M_{\rm d}= 0.001 M_\odot$ and $e_{\rm b}=0.75$ from  Equation ({\ref{allim}}) that $\alpha \sim 10^{-4} \sin^2{i}$ for p=1.
Such levels of turbulence are extremely low.
Also such thin layers suggest that the density of dust
near the disc midplane is greater than the gas density.
This configuration is subject to various instabilities,
such as shear instability and streaming instability \citep{Youdin2002,Youdin2005}.
It is not clear that such thin layers
can exist.

Less extreme values of $\alpha$
can occur if the occulting material resides
at smaller radii, so that $T/H$ is larger. The smaller radii could occur due to the inward drift of solids.
The velocity constraint in Equation (\ref{vobs}) can be satisfied by the occulting solids because the precession
rate is controlled by the more extended gas disc.
We consider the case that $p=1$ and apply the rigid tilt approximation
that assumes the disc remains flat during its evolution.
In that case, the velocity
constraint is satisfied provided that
the outer radius of the gaseous disc satisfies
\begin{equation}
r_{\rm out} \simeq 0.34 \left(\frac{r_{\rm s}/{\rm AU}}{v/{\rm (AU/yr)}}
\right)^{2/3} {\rm AU},
\label{routs}
\end{equation}
where $r_{\rm s}$ is the radius of the occulting solids and $v$ is given by Equation (\ref{vobs}). This equation holds for $r_{\rm in}=1 {\rm AU} \ll r_{\rm s} < r_{\rm out}$.
If the occulting occurs
at $r_{\rm s}=4$ AU for the $p=1$ disc model described in the previous paragraph with $
M_{\rm d}=0.001 M_\odot$, then $T/H \sim 0.5 \sin{i}$ at $r=r_{\rm s}$, then $\alpha \sim 10^{-4} \sin^2{i}$ by Equation ({\ref{allim}}), and  
$r_{\rm out}= 16$ AU by Equation (\ref{routs}).
Higher values of $\alpha \ga 0.01$ can occur
for very small disc masses $M_{\rm d} \la 10^{-5} \sin^2{i} \, M_\odot$.
For comparison, in the case of HL Tau, \cite{Pinte2016} found that a thin sublayer of millimeter
sized grains $T/H \la 0.2$ could account for the observed properties of the system
that in turn imposed an upper limit on $\alpha \sim 3 \times 10^{-4}$.

 The Stokes number for dust grains compares the stopping time $t_{\rm s}$ to the dynamical time. %Significant radial drift occurs when $\rm Stk \sim 0.1 - 10$.  
For a disc with $p=1$, its value at radius $r$ is estimated as
\begin{equation}
    {\rm Stk} = \Omega \, t_{\rm s} \simeq 6 \times 10^{-4} \bigg( \frac{r_{\rm g}}{50\, \rm \micro m}\bigg) \bigg( \frac{M_{\rm d}}{0.001\, \rm M_{\odot}}\bigg)^{-1} \bigg( \frac{r_{\rm out}}{16\, \rm AU}\bigg) \bigg( \frac{r}{4\, \rm AU}\bigg)
\end{equation}
\citep[cf.][]{Fromang2009}, where $r_{\rm g}$ is the grain size. With $r_{\rm g} = 50\, \rm \micro m$, $M_{\rm d} = 0.001\, \rm M_{\odot}$, $r_{\rm out} = 16\, \rm AU$, and $r = r_{\rm s} = 4\, \rm AU$, then ${\rm Stk} \simeq 6\times10^{-4}$. For these parameters, the dust is well coupled to the gas. The inward radial drift velocity due to gas drag is $v_{\rm r} \sim (H/r)^{2} \, Stk \, \Omega r \sim 10^{-5} \Omega r$ with $H/r = 0.1$ and $r = 16 \, \rm AU$ \citep{Armitage2013}.  The drift timescale near the disc outer edge is then of order $10^{6}$ years. Its numerical value in this case is not sensitive to $p$ for $0.5 \le p \le 1.5$ Shorter drift timescales occur for a less massive disc.

 Disc warping could also influence the effective value of $T$ by making the requirements on the thickness on the solids layer even stronger (thinner layer), but we do not consider its effects here.
%For discs that satisfy the velocity constraint, the warping is  small,
%even though the discs are broad. This occurs because the velocity constraint
%imposes a limit on the disc precession period that implies it is longer
%than the sound crossing timescale.
Another possibility is that the disc does not contain significant amounts of
gas with associated turbulence, but instead essentially consists of only solids. The coherence of the disc or ring
against the effects of differential precession is due to the self-gravity
\citep{Chiang2004}.  For such a ring, some of the linear theory results in this paper still hold, such as those in Figures \ref{ivrst},  \ref{fig:imax}, and \ref{fig:er}.

\section{Summary}
\label{concs}

We have analysed the behavior of a mildly tilted low mass  circumbinary disc in an eccentric orbit binary star systems by means of SPH simulations and linear theory.
The disc undergoes nonuniform precession and tilt oscillations due to the effects of the binary eccentricity (e.g., Figs.~\ref{e08_1} and \ref{precrate}). For moderately broad discs (whose outer radii are a few times the inner radii)
with typical protostellar disc
parameters, the disc can precess coherently with little warping. Larger discs can undergo breaking (Fig.~\ref{SPH2}).
For small initial tilts, the results of the SPH simulations agree well with linear theory (e.g., Fig. ~\ref{fig:imax}).
The amplitude of the tilt oscillations increases with binary eccentricity. The disc tilt undergoes damped oscillation in time and ultimately approaches a coplanar alignment with the binary.

We have analyzed a model for 
binary KH 15D that is based on a mildly tilted precessing disc that orbits an eccentric binary.
The model suggests that the disc tilt relative to the binary orbit is currently at a minimum value and that the retrogade precession rate is currently at its largest value.
We considered a period constrained model for the disc, along the lines of the previous circular orbit binary studies  \citep{Lodato2013,Foucart2014},
but taking into account the binary eccentricity. We find that the large binary eccentricity changes the inferred disc outer radii by a small amount.
To satisfy the disc tilt lifetime
requirements, the disc $\alpha$
value must be small, less than about 0.001, as is also consistent with the earlier studies.

We then considered a model in which
the outer disc edge precession velocity is constrained by the observed changes in the binary eclipse properties \citep[e.g.,][]{Aronow2018}.
We determined the relation between the disc inner and outer radii subject to this constraint. We find that discs
whose inner radius is tidally truncated by the binary
typically have outer radii of $\sim 30 -170$ AU depending on the disc density profile. The disc outer radii are reduced if there is inward radial migration of solids that are 
responsible for the binary occultation. 
Narrow disc radii are about $3$ AU,
in agreement with \cite{Chiang2004}. 
%To satisfly the tilt lifetime requirements in this case, the disc $\alpha$ value need not be very small, values of $\sim \alpha=0.01$ are adequate.

The recent reappearance of Star B places strong constraints on the thickness of an occulting layer of solids/dust. 
The most reasonable models involve a thin layer of 
dust that has settled towards the midplane of a low mass
gaseous disc $M_{\rm d} < 0.001 M_\odot$
and has migrated considerably inward.
Such thin layers suggest that the 
 disc turbulence is very weak  $\alpha \ll 0.001$.
Stronger turbulence can occur for smaller mass discs.
For a narrow ring, less extreme settling and levels of turbulence are required. Another possibility is that the disc/ring consists of a thin disc of solids with little gas \citep[e.g.,][]{Chiang2004}.

As noted in \cite{Martin2017}, it is also possible that the disc is instead evolving to a polar (perpendicular) alignment with the binary. For this to occur, the disc tilt needs to be 
$i \ga 30^\degree$.

\section*{Acknowledgments} 
 We much appreciate Hossam Aly for beneficial conversations and for carefully reviewing the paper. We thank Daniel Price for providing the {\sc phantom} code for SPH simulations and acknowledge the use of SPLASH \citep{Price2007} for the rendering of the figures. SL thanks Eugene Chiang for insightful discussions. We acknowledge support from NASA through grant NNX17AB96G.  Computer support was provided by UNLV's National Supercomputing Center.

\bibliographystyle{mnras}
%\bibliography{smallwood} % if your bibtex file is called example.bib
\bibliography{main}

% Don't change these lines
\bsp	% typesetting comment
\label{lastpage}
\end{document}